\documentclass[aps, pre, twocolumn]{revtex4-2}
\usepackage[dvips]{graphicx}
\usepackage{amsmath}
\usepackage{bm}
\usepackage{color}
\usepackage{natbib}
\usepackage{comment}
\usepackage[utf8]{inputenc}
\usepackage{MnSymbol}

\begin{document}
\title{Motion Ordering in Cellular Polar--polar and Polar--nonpolar Interactions}

\author{Katsuyoshi Matsushita}
\affiliation{Department of Mathematical and Life Sciences, Hiroshima University, Higashi-Hiroshima, Hiroshima, 739-0046, Japan}
\author{Taiko Arakaki}
\affiliation{Department of Mathematical and Life Sciences, Hiroshima University, Higashi-Hiroshima, Hiroshima, 739-0046, Japan}
\affiliation{Department of Biological Science, Osaka University, Toyonaka, Osaka, 560-0043, Japan} 
\author{Koichi Fujimoto}
\affiliation{Department of Mathematical and Life Sciences, Hiroshima University, Higashi-Hiroshima, Hiroshima, 739-0046, Japan} 

\begin{abstract}
We examine the difference in motion ordering between cellular systems with and without information transfer to evaluate the effect of the polar--polar interaction through mutual guiding, which enables cells to inform other cells of their moving directions. We compare this interaction with the polar--nonpolar interaction through cell motion triggered by cellular contact, which cannot provide information on the moving directions. We model these interactions on the basis of the cellular Potts model. We calculate the order parameter of the polar direction in the interactions and examine the cell concentration and surface tension conditions of ordering. The results suggest that the polar--polar interaction through mutual guiding efficiently induces the motion ordering in comparison with the polar-nonpolar interaction for contact triggering, except in cases of weak driving. The results also show that the polar--polar interaction efficiently accelerates the collective motion compared with the polar--nonpolar interaction.
\end{abstract}


\maketitle

\section{Introduction}
The collective motion of eukaryotic cells is indispensable in various developmental and pathological phenomena \citep{Rorth:2009, Friedl:2009, Weijer:2009}. In many cases, cells stabilize their motion order.
Various biological stabilization mechanisms have been proposed \citep{Haeger:2015, Coates:2001, Stramer:2016} from active matter and biological physics perspectives. These proposals promote our understanding of collective cell motion \citep{Marchetti:2013, Hakim:2017}. The mechanisms utilize various factors such as the transmission of chemical waves \citep{Swaney:2010, Camley:2016, Camley:2018}, the contact following \citep{Fujimori:2019, Hiraiwa:2020, Hayakawa:2020} or inhibition of locomotion\citep{Schnyder:2017, Hiraiwa:2019, Jain:2022}, leader cells\citep{Omelchenko:2003, Haga:2005, Trepat:2009, Kabla:2012},  stress waves \citep{Serra-Picamal:2012, Notbohm:2016, Yabunaka:2017a, Yabunaka:2017b, Tlili:2018, Fukuyama:2020}, cell cycle \citep{Li:2021, Alhashem:2022}, the persistence of motion \citep{Szabo:2006, Matsushita:2019, Matsushita:2020a}, cytoskeleton remodeling \citep{Sato:2015a, Sato:2015b}, substrate stiffness \citep{Lo:2000, Sunyer:2016, Pallares:2023}, and guiding using the cell--cell adhesion polarity or surface tension gradient \citep{Matsushita:2018, Okuda:2022a, Matsushita:2022a, Okuda:2022b}. The roles of these factors in the ordering mechanism of collective motion remain unclear, especially from the physics viewpoint.

\begin{figure}[t]
    \begin{center}
        \includegraphics[width=0.7\linewidth]{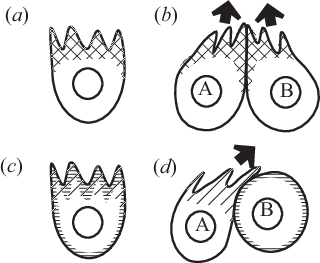}
        \caption{(a) Model cell for mutual guiding. The cross-hatched region contains a high concentration of adhesion molecules. This high concentration is origin of polar characteristics, namely polarity. The sawtooth shape expresses the leading edge of a cell. (b) Schematic view of the collective propulsion due to adhesion. The adhesion between two cells, A and B, in their leading edges, extends them in the directions of their leading edges and result in their propulsion. The arrows represent the direction of motions driven by the propulsion. This propulsion is based on the polar--polar interaction. (c) Model cell for contact--triggering. The slashed region contains a high concentration of a receptor, and the border region contains a high concentration of the corresponding ligand. (d) Cell A moves upon being triggered by the ligand in cell B. In this panel, the receptor and ligand are only illustrated for cells A and B, respectively. As shown by the slashed and border regions, all cells commonly have localized receptor concentration only in their leading edges and have uniform ligand concentration in the whole of edges. In this sense, the propulsion is based on the polar--nonpolar interaction. Note that both cells A and B actually move in the model, although the arrow expressing motion is illustrated only for cell A for simplicity.}\label{fig:models}
    \end{center}
\end{figure}
The motion--ordering efficiency of these factors gives a helpful hint to this ordering mechanism. 
Here, we define efficiency as the achievement of higher--order cellular motions with increasing strength of the factors. Namely, using the order parameter of motion directions, $P$, as a function of the strength $E$, we can simply define the efficiency $e$ as
\begin{align}
e = \frac{\partial P}{\partial E}.
\end{align}
The motion--ordering efficiency cannot be easily examined in experiments, even in vitro, because of the difficulty in the experimental control of the factors. 
As an alternative to experimental examination, by theoretical examination, one can easily control the individual cell factors and consider the pure effects of these factors reductively \citep{Alert:2020}. 
Furthermore, the theoretical examination based on a physical model enables a simple thought experiment to evaluate the motion--ordering efficiency and provide physical insights into collective motion. 
Using these merits of theoretical examination, we shed light on the polar--polar interaction using cell--cell adhesion \citep{Coates:2001,Matsushita:2018} as a model case and aim to evaluate its motion--ordering efficiency. 

A model cell for the polar--polar interaction through adhesion is illustrated in Fig.~\ref{fig:models}(a). Polar adhesion, which is simply called polarity, induces a surface tension gradient from the rear to the leading edges in the cell-cell contact region in Fig.~\ref{fig:models}(b). The gradient drives the extension of the leading edges  of cells in contact, and the membrane in the edges flows to the rear edges by the Marangoni effect, similarly to self-propelled droplets \citep{Levan:1981}. The gradient drives the collective motion of cells \citep{Matsushita:2018, Shellard:2019, Matsushita:2022a, Okuda:2022a, Itatani:2022, Yadav:2022, Pajic-Lijakovic:2023, Sato:2023}. We call this mechanism ``mutual guiding''.

The polar--polar interaction through cell--cell adhesion was  previously proposed on the basis of experiments using {\it Dictyostelium discoideum} cells \citep{Sesaki:1996, Coates:2001} and theoretically shown to promote their motion ordering \citep{Matsushita:2018} with characteristic alignments \citep{Matsushita:2017}. Additionally, theoretical examinations showed that the ordering arises as the effect of the cellular surface-tension gradient in tissue \citep{Matsushita:2020a,Matsushita:2022a}. In this case,  the tension gradient reflects the motion feedback \citep{Matsushita:2019, Matsushita:2021a} similarly to self-propelled disks \citep{Weber:2013,Hanke:2013,Hiraoka:2016,Hiraoka:2017}. This motion feedback promotes the ordering. The efficiency of mutual guiding has not been sufficiently examined in contrast to the self-propelled disks \citep{Matsushita:2019}. We aim to clarify this efficiency {and its origin through this work.

To design the clarifying method of the origin of the efficiency, we focus on the characteristic of the polar--polar interaction: cells can inform other cells of the direction of their leading edges through polar adhesion. In fact, the adhesion exerts itself mainly on the leading edges of cells and thereby guides the leading edge extension of other interacting cells only in the direction of the leading edges. Therefore, we can hypothesize that the efficiency of mutual guiding originates from the information transfer in the leading edge direction between cells. This hypothesis leads us to the expectation that the efficiency measurably diminishes when the information transfer based on the polar characteristic is insufficient. Therefore, a natural way to examine the possibility of information transfer as the origin of the efficiency is to compare the cells with mutual guiding to those without this information transfer in their interaction.

For this purpose, we consider comparing mutual guiding with a hypothetical mechanism with partial loss of polarity. With such a hypothetical mechanism, we introduce ``contact--triggering'', which is analogous to the contact attraction of locomotion \citep{Hiraiwa:2020} in the sense that cellular contacts drive the motion. In this case, we assume the polarized concentration of the receptor molecule and the uniform concentration of its ligand as shown in Fig.~\ref{fig:models}(c). In addition, the receptor responding to its ligand through cellular contact is assumed to trigger the leading edge extension of the cell by reducing its surface tension, as shown in Fig.~\ref{fig:models}(d).
Owing to the uniform concentration of the ligand, this system partially loses polarity. In this sense, the cellular interaction through contact--triggering is the polar--nonpolar interaction. 

In this system, we emphasize that cells cannot inform other cells of their polarity directions. This is because the ligand concentration is uniform. 
Even without polarity in the ligand concentration, contact--triggering can stabilize the motion order cooperatively, similarly to mutual guiding \citep{Matsushita:2023a, Matsumoto:2024}. 
Contact--triggering is in contrast to mutual guiding in terms of the information transfer between cells. The polarity of the ligand in mutual guiding exists as shown in Fig.~\ref{fig:models}(b). Therefore, cells can mutually inform the direction of their polarity. 

Note that this difference between the two systems enables us to compare the effects of different symmetries. In the case of mutual guiding, the adhesion concentration has no inversion symmetry, and the interaction of cells through their contacts becomes a polar--polar type. In contrast, in the case of contact--triggering, the ligand concentration has a symmetry against inversion, and the receptor concentration has no inversion symmetry. Therefore, the interaction through their contacts becomes a polar--nonpolar type. The difference in symmetry may result in their collective behavior. This examination may clarify the effects of this symmetry difference from the physics viewpoint.

In this work, we consider model cells with these interactions based on the two-dimensional cellular Potts model. We compare these cells in terms of the order parameter of polarity and examine the ease of ordering of suspended or aggregated cells in an extracellular matrix (ECM) as adhesive or inadhesive scaffolds, respectively \citep{Hawkes:1982, Mecham:2011}. We control the cell density and interface tension between cells and ECM, and calculate the order parameter.
We numerically show that mutual guiding with the polar--polar symmetry enhances the motion ordering efficiency compared with contact--triggering under strong driving with the polar--nonpolar symmetry. Namely, the former efficiency $dP/d\varepsilon$ at the onset of motion ordering is larger than the latter efficiency $dP/d\delta$, where $\varepsilon$ and $\delta$ are the strengths of mutual guiding and contact--triggering, respectively.



\section{Models}

We use the two-dimensional cellular Potts model \citep{Graner:1992,Glazier:1993, Graner:1993}, which has been widely used for expressing multicellular processes \citep{Anderson:2007, Scianna:2013, Hirashima:2017} because of its ease of dealing with cellular interactions through interface tension. This ease is suitable for our purpose, which is the evaluation of the surface tension gradient in motion ordering.
In contrast, because of the discreteness of the model on a lattice, the accuracy of surface tension is limited by the representation capability of the discrete lattice.
In this model, the Potts state $m(\bm r)$ at the lattice site $\bm r$ is defined on a square lattice with the linear size $L$. The state expresses the occupation cell index at the site. Here, the cell index takes a number in ${1, 2, \dots, N}$, and $N$ is the number of cells. The set of $m(\bm r)$, $\{m(\bm r)\}$ for all lattice sites expresses the cellular configuration. The present model expresses ECM with $m(\bm r)$ = 0. 

The configuration of this Potts state is stochastically updated with a Monte Carlo method and simulates a time course of cellular configurations. The Monte Carlo method consists of a consecutive series of Monte Carlo steps. Each Monte Carlo step conventionally consists of 16$L^2$ trials of a state copy \citep{Graner:1992}. The state copy is a copy to the randomly chosen site $\bm r$ from the randomly selected neighbor $\bm r'$, similarly to the voter model \citep{Clifford:1973,Liggett:1985}. Here, the set of neighboring sites for $\bm r$ is defined by the nearest and next--nearest sites. Then, the copy is accepted with the Metropolis probability $P$ = $\min \{ 1, P(\{m'(\bm r)\})/P(\{m(\bm r)\})\}$. This consecutive update procedure corresponds to a Metropolis--Hastings update \citep{Landau:2021} using the voter model as a proposed copy. Here, ${m'(\bm r)}$ expresses the state after the copy from $\bm r$ to $\bm r'$. $P(\{m(\bm r)\})$ denotes the occurrence probability of the state $\{m(\bm r)\}$ and is proportional to the form of the Boltzmann factor, $\exp\left[-\beta {\cal H}(\{m(\bm r)\})\right]$, where $\beta$ is the strength of cell shape fluctuation and ${\cal H}(\{m(\bm r)\})$ denotes the energy of the state $\{m(\bm r)\}$.

The energy ${\cal H}(\{m(\bm r)\})$ has four terms:
\begin{align}
{\cal H}(\{m(\bm r)\}) &= {\cal H}_{\rm C}(\{m(\bm r)\}) + {\cal H}_{\rm E}(\{m(\bm r)\}) \nonumber \\ &+ {\cal H}_{\rm A}(\{m(\bm r)\}) + {\cal H}_{\rm P}(\{m(\bm r)\}).
\end{align}
The first term denotes the surface free energy between cells. The second term represents the surface free energy between cells and ECM. The third term denotes the area elastic energy. The fourth term denotes the propulsion term of mutual guiding or contact--triggering.
 
The first term takes the form
\begin{align}
{\cal H}_{\rm C}(\{m(\bm r)\}) = \gamma_{\rm C}\sum_{\bm r\bm r'} \eta_{m(\bm r)m(\bm r')}\eta_{m(\bm r)0}\eta_{m(\bm r')0}.
\end{align}
Here, $\eta_{mm'}$ = $\left[1-\delta_{mm'}\right]$ and $\delta_{mm'}$ is Kronecker's delta. The second term takes the form. 
\begin{align}
{\cal H}_{\rm E}(\{m(\bm r)\}) &= \gamma_{\rm E}\sum_{\bm r\bm r'}\eta_{m(\bm r)m(\bm r')}\nonumber \\ &\times \left[\eta_{m(\bm r)0}\delta_{m(\bm r')0} + \eta_{m(\bm r')0}\delta_{m(\bm r)0}\right].
\end{align}
Here, $\gamma_{\rm C}$ and $\gamma_{\rm E}$ denote the surface tension between cells and that between cells and ECM, respectively.
For $\gamma_{\rm C}$ $>$ $2\gamma_{\rm E}$, cells are suspended in ECM \citep{Glazier:1993}, otherwise, they aggregate. Therefore, $\gamma_{\rm E}$ =  $\gamma_{\rm C}/2$ gives the transition condition between the suspension and aggregation states.

The third term takes 
\begin{align}
{\cal H}_{\rm A}(\{m(\bm r)\}) = \kappa A \sum_m \left[ 1 - \frac{\sum_{\bm r}\delta_{mm(\bm r)} }{A}\right]^2. \label{eq:propulsion}
\end{align}
Here, $\kappa$ denotes the elastic modulus for the cell area, and $A$ represents the reference area of the cells. 

The fourth term takes \citep{Matsushita:2017}
\begin{align}
{\cal H}_{P} = \sum_{ab}\gamma_{a b}\sum_{\bm r \bm r'} \eta_{m(\bm r)m(\bm r')} \left[1-\frac{\rho_{a}(\bm r)\rho_b(\bm r')}{\rho_a^0\rho_b^0}\right]. \label{eq:H_P}
\end{align}
Here, $a$ and $b$ are the indexes for the type of molecule. $\rho_{a}(\bm r)$ denotes the molecular concentration of $a$ at $\bm r$, and $\rho_a^0$ is the corresponding saturation concentration. $\gamma_{ab}$ denotes the reduction in the amplitude of the surface tension caused by the binding between molecules $a$ and $b$. If the origin of the reduction is the normal force on the leading edge in the case of contact--triggering, the surface tension is effectively recognized as the normal force multiplied by the surface curvature radius, as in the Laplace law \citep{DeGennes:2004,Safran:2013}. 

In the case of mutual guiding, we assume that molecules $a$ and $b$ are the same as the cell--cell adhesion molecule, namely, a homophilic adhesion molecule \citep{Takeichi:2014}. We additionally consider the anisotropy of its concentration \citep{Sesaki:1996,Coates:2001,Zajac:2002,Vroomans:2015,Matsushita:2017}. The concentration $\rho_{a}(\bm r)$ has the following first-order term of the multipole expansion: \citep{Matsushita:2017}
\begin{align}
\rho_{a}(\bm r) &= \frac{\bar \rho_{\rm ad}}{2} q_m(\bm r) \eta_{m(\bm r)0},\label{eq:concentration_polar}\\
q_m(\bm r) &= \left[1+\bm \mu_m \cdot \bm e_{m(\bm r)}(\bm r)\right].
\end{align}
Here, $\bar \rho_{\rm ad}$ is the highest concentration on a cell and $\bm e_m(\bm r)$ is the unit vector indicating $\bm r$ from the center of the $m$th cell $\bm R_m$. The unit vector $\bm \mu_m$ is the so-called cell polarity, and in this case, we define the polarity from the concentration polarity of cell-cell adhesion. We also assume that the cell polarity takes the same direction as the cell's leading edge. We further assume that $\bm \mu_m$ and $\bm R_m$ are slow variables, and we define their dynamics at the end of the explanation for this term. 

By substituting $\rho_{a}$ and $\rho_b$ with the expression in Eq.~\eqref{eq:concentration_polar}, we obtain the part of mutual guiding in the summation for $a$ and $b$ in eq.~\eqref{eq:H_P} as \citep{Matsushita:2018} 
\begin{align}
-\varepsilon  \sum_{\bm r \bm r'} \eta_{m(\bm r)m(\bm r')} \eta_{m(\bm r)0}\eta_{m(\bm r')0} \left[q_{m(\bm r)}(\bm r)q_{m(\bm r')}(\bm r') \right]+ {\rm const}. \label{eq:mutual_guiding}
\end{align}
Here, $\varepsilon$ = $\gamma_{\rm ad}(\bar\rho_{\rm ad}/\rho_{\rm ad}^0)^2/4$ is the degree of tension reduction in the case of mutual guiding. For this mutual guiding, $\gamma_{\rm ad}$ and $\rho_{\rm ad}^0$ denote the surface tension reduction and the corresponding highest molecule concentration, respectively. The subscript ${\rm ad}$ indicates the molecule indexes $a$ and $b$ for the adhesion molecule. The constant part is renormalizable to ${\cal H}_{\rm C}$ and, therefore, can be omitted hereafter. 

Similarly to mutual guiding, we can consider the concentrations of different molecules, namely, the receptor and ligand, for $\rho_a$ and $\rho_b$ in the case of contact--triggering. In this case, the concentration of a ligand is uniform and that of the receptor is polar. Therefore, we can take 
\begin{align}
\rho_{a}(\bm r) &= 
\begin{cases}
& \rho_{\rm lig}(\bm r) = \bar \rho_{\rm lig} \eta_{m(\bm r)0}, \\
& \rho_{\rm rec}(\bm r)  = \frac{\bar \rho_{\rm rec}}{2} \tilde{q}_m(\bm r) \eta_{m(\bm r)0},
\end{cases}\\
\tilde{q}_m(\bm r) &= \left[1+{\bm \nu}_m \cdot \bm e_{m(\bm r)}(\bm r)\right]
\end{align}
as the concentrations of ligand and receptor molecules, respectively. This formula is analogous to the case of the tissue interface \citep{Matsushita:2020a, Matsushita:2022a,Pajic-Lijakovic:2023}. By substituting $\rho_{a}$ and $\rho_{b}$ in Eq.~\eqref{eq:concentration_polar} with these expressions, we obtain the contribution of contact--triggering in the summation for $a$ and $b$ in Eq.~\eqref{eq:H_P} as 
\begin{align}
    -\delta\sum_{\bm r \bm r'} \eta_{m(\bm r)m(\bm r')} \eta_{m(\bm r)0}\eta_{m(\bm r')0}\left[\tilde{q}_m(\bm r)+\tilde{q}_m(\bm r')\right] + {\rm const }. \label{eq:contact-triggering}
\end{align}
Here, $\delta$ = $\gamma_{\rm lr}(\bar\rho_{\rm rec}/\rho_{\rm rec}^0)(\bar\rho_{\rm lig}/\rho_{\rm lig}^0)/2$  is the degree of tension reduction for contact-triggering. In this case, $\gamma_{\rm lr}$, $\rho_{\rm rec}^0$, and $\rho_{\rm lig}^0$ denote the surface tension reduction and the saturation concentrations of receptor and ligand, respectively. The subscripts $\rm lig$, $\rm rec$, and ${\rm lr}$ denote the molecule indexes $a$ and $b$ for the ligand, receptor, and both molecules, respectively.   ${\bar \rho}_{\rm rec}$ and ${\bar \rho}_{\rm lig}$ are the highest values of the corresponding concentrations $\rho_{\rm rec}$ and $\rho_{\rm lig}$, respectively. 
The unit vector ${\bm \nu}_m$ is also cell polarity defined by the concentration polarization of the receptor. We assume that the direction of ${\bm \nu}_m$ agrees with the leading edge of cells. As mentioned above, ${\bm \mu}_m$ also takes the direction of the leading edge. Therefore, these are the two polarities of ${\bm \mu}_m$ = $\bm \nu_m$. Here, by defining the unified direction of cell polarities as the direction of the leading edge (${\bm p}_m$), we can impose $\bm p_m$ = ${\bm \mu}_m$ = ${\bm \nu}_m$. Hereinafter, we use $\bm p_m$ instead of ${\bm \mu}_m$ and ${\bm \nu}_m$ for simplicity.

For these mechanisms, the range of contributions in Eq.~\eqref{eq:mutual_guiding} is from 0 to 4$\varepsilon$ and that in Eq.~\eqref{eq:contact-triggering} is from 0 to 4$\delta$. These ranges agree with each other for $\varepsilon$ = $\delta$. In this sense, we quantitatively compare the difference between these mechanisms under the supposition that the contributions agree with each other for $\varepsilon$ and $\delta$.

The fourth term depends on the dynamics of the slow variables $\bm p_m$ and $\bm R_m$. The dynamics of $\bm p_m$ obeys the equation of motion \citep{Szabo:2006,Kabla:2012, Matsushita:2019}
\begin{align}
\frac{d \bm p_m }{dt} =  \frac{1}{c\chi} \left[\frac{d \bm R_m}{dt} - \left(\bm p_m \cdot \frac{d \bm R_m}{dt}\right) \bm p_m \right].\label{eq:p}
\end{align}
Here, $c$ and $\chi$ denote the lattice constant and the ratio of the relaxation time of $\bm p_m$ to that of $\bm R_m$, respectively. This equation is integrated once by the Euler method and, simultaneously, $\bm R_m$ is set at the 
 center of the cell mass $\sum _{\bm r}\delta_{mm(\bm r)}{\bm r}/\sum_{\bm r}\delta_{mm(\bm r)}$. Note that under the setting of the norm $|\bm p_m|$ = 1, Eq.~\eqref{eq:p} does not change the norm of $\bm p_m$.

\begin{figure*}[t]
    \begin{center}
        \includegraphics[width=0.8\linewidth]{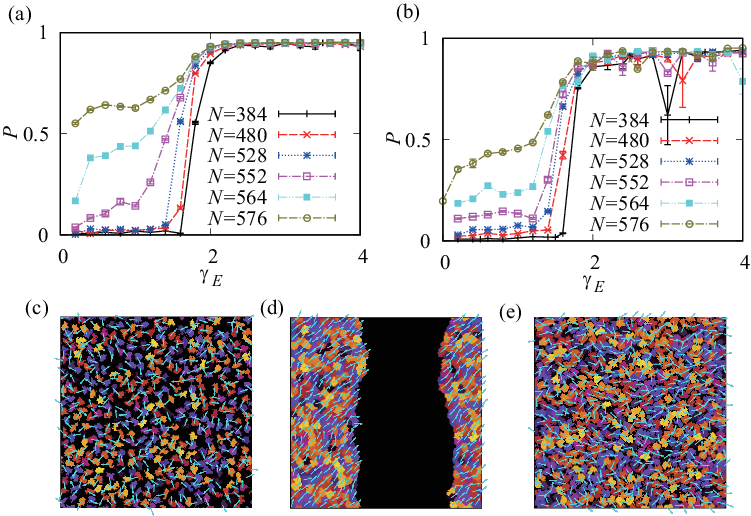}
        \caption{Order parameters as functions of $N$ and $\gamma_E$ for (a) mutual guiding ($\varepsilon$ = 0.2) and (b) contact--triggering ($\delta$ = 0.2). Snapshots of cell configurations: (c) mutual guiding with $N$ = 384 and $\gamma_{\rm E}$ = 1.0, (d) mutual guiding with $N$ = 384 and $\gamma_{\rm E}$ = 4.0, and (e) mutual guiding with $N$ = 576 and $\gamma_{\rm E}$ = 1.0. In these snapshots, the different colors of domains of Potts states indicate different cells. The arrow on each cell represents the polarity direction of the cell.}\label{fig:order_parameter}
    \end{center}
\end{figure*}
 We employ the system size of $L$ = 192 for empirically tractable simulation. To examine the total area fraction of cell occupation $\phi$ up to 1, we choose $N$ = 384, 480, 528, 552, 564, and 574. We employ $A$ = 64. Hence, the corresponding $\phi$ values ($\phi$ = $VN/L^2$) for the values of $N$ are respectively about 67, 83, 92, 95, 98, and 100\%. These $\phi$ values enable us to observe the transition from individual to collective motion with increasing $\phi$ at a low $\gamma_E$. The parameters that can control the deformation ability of cells, $\gamma_{\rm C}$,  $\kappa$, and $\beta$ are set to $4.0$, $1.0$, and $0.2$, respectively, because they can easily enable cells to deform for their motions. In the case of the $\gamma_{\rm C}$ value, cells aggregate at values of $\gamma_{E}$ $>$ 2.0 \citep{Glazier:1993}. We employ $c$ as a unity. We choose $\chi$ of $5.0$ empirically for the ease of ordering for these parameters \citep{Matsushita:2021a} and its change does not affect the result when the value is larger than 2.0. For $\delta$ and $\varepsilon$, we first use 0.2 to measure the effects of $\gamma_{\rm E}$ and $\phi$ and then vary them to examine the effects of these driving terms. 

 To observe the ordering of cell motion, we consider the order parameter of polarity $\bm p_m$,
 \begin{align}
  P = \frac{1}{NT}\left|\int_{t_0}^{t_0+T} \sum_m \bm p_m(t)\right|,
 \end{align}
 and the collective velocity
 \begin{align}
  v = \frac{1}{NT}\left|\int_{t_0}^{t_0+T} \sum_m \bm d_m(t)\right|.
 \end{align}
 Here, $\bm d_m(t)$ = $\bm R_m(t)$ $-$ $\bm R_m (t-1)$ is the displacement of the $m$th cell per Monte Carlo step.
 To calculate these values of steady states, we first relax the state during $t_0$ = 10$^5$ Monte Carlo steps and integrate it during the time period $T$ = 10$^5$ Monte Carlo steps. $P$ and $v$ in the steady states do not depend on the initial state, where the cells expressed by the domain of $ m(\bm r)$ are aligned in a lattice array and ${\bm p}_m$ takes a random direction. \citep{Matsushita:2017}

 \section{Results}

 Our examination starts with the determination of the ordering condition of motion. The ordering results from the cell-cell interaction through intercellular contacts. The contacts frequently occur at high cell densities. Therefore, the ordering condition is expected to strongly depend on the number of cells $N$ per system size $L^2$ (or the area fraction $\phi$). This effect is especially prominent when the cells prefer to contact ECM rather than each other because cells cannot spontaneously make contact with each other. In addition, the contact formation between cells depends on their surface tension. In particular, cell aggregation is also expected to affect the ordering of cells through intercellular contacts. The cell aggregation is stabilized at a value of $\gamma_{\rm E}$ $>$ $\gamma_{\rm C}/2$ = 2.0 in our setting. Therefore, the ordering condition changes at $\gamma_{\rm E}$ = 2.0. To confirm these expectations, we calculate $P$ as a function of $N$ and $\gamma_{\rm E}$ and examine the difference in motion ordering between the two mechanisms.

\begin{figure*}[t]
    \begin{center}
        \includegraphics[width=0.85\linewidth]{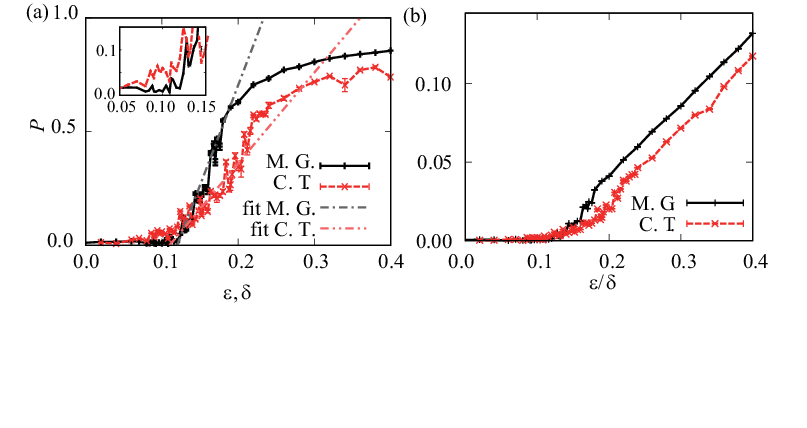}
        \caption{(a) Order parameter $P$ and (b) collective velocity $v$ as functions of $\varepsilon$ for mutual guiding (M. G.) and $\delta$ for contact--triggering (C. T.) at $\gamma_{\rm E}$ = 1.0 and $N$ = 576 ($\phi$ = 100\%). 
         In (a), the inset shows the magnified version of the figure in the range from $\varepsilon$ = $\delta$ = 0.05 to 0.15. The dashed-dotted and dashed-double-dotted lines represent the fitting lines for calculating the efficiencies of $dP/d\varepsilon$ and $dP/d\delta$, respectively.
         }\label{fig:comp_suspension}
    \end{center}
\end{figure*}

 Figure~\ref{fig:order_parameter}(a) shows the order parameters for mutual guiding at $\varepsilon = 0.2$ and Fig.~\ref{fig:order_parameter}(b) shows those for contact--triggering at $\delta = 0.2$. At the low area fraction of $N$ = 384 ($\phi$ = 67\%), $P$ is zero at a low $\gamma_{\rm E}$ and is about unity at a high $\gamma_{\rm E}$, commonly for both mutual guiding and contact--triggering. Thus, the motion ordering only occurs at a large $\gamma_{\rm E}$. This result suggests that the aggregation at a large $\gamma_{\rm E}$ promotes motion ordering. In fact, the absence of aggregation at $\gamma_{\rm E} = 1.0$ is observed in the cell configuration in Fig.~\ref{fig:order_parameter}(c), and the presence of aggregation at $\gamma_{\rm E} = 4.0$ is observed in that in Fig.~\ref{fig:order_parameter}(d).
 The transition from disordered to ordered states occurs abruptly at $\gamma_{\rm E}$ slightly smaller than 2.0, which is expected from above. 
 
 As $N$ (or $\phi$) increases, $P$ at a low $\gamma_E$ increases from $N=526$ ($\phi$ = 92\%). $P$ at $N$ = 576 ($\phi$ = 100\%) reaches around 0.5 even at $\gamma_{\rm E}$ $<$ 2.0. This is observable from the ordering direction of polarity on cells in Fig.~\ref{fig:order_parameter}(e). Therefore, cell condensation at a large $\phi$ also induces the ordering of cell motion.

 The quantitative difference in these results between mutual guiding and contact--triggering appears at a low $\gamma_{\rm E}$ and around $\phi$ $\gtrsim$ 92\%. Therefore, to compare these mechanisms in terms of motion--ordering efficiency,  we focus on the case of low $\gamma_{\rm E}$
and $N$ = 576 ($\phi$ = 100\%) and evaluate the dependence of $P$ on the driving force strengths of $\varepsilon$ and $\delta$.

Figure~\ref{fig:comp_suspension}(a) shows $P$ as a function of $\varepsilon$ for mutual guiding and $\delta$ for contact--triggering. For mutual guiding, $P$ at a low $\varepsilon$ is almost zero, which therefore indicates a disordered state. The disordered state is stable even as $\varepsilon$ increases to 0.1. The disordered state is an effect of the inhibition of the cell-cell interaction by ECM. $P$ starts to increase above $\varepsilon$ = 0.1 and then increases with $\varepsilon$. Therefore, we can expect that the interaction will overcome the inhibition due to ECM through $\varepsilon$ above a threshold $\varepsilon_c$ of 0.1. 
The increase in $P$ decelerates around $\varepsilon$ = 0.3. For contact--triggering, $P$ and, therefore, the state of motion ordering take a similar behavior as a function of $\delta$. One exception point is that the onset of the increase in $P$ is below 0.1, as shown in the inset of Fig.~\ref{fig:comp_suspension}(a). Mutual guiding is inferior in motion--ordering efficiency for at a low driving force compared with contact--triggering, that is, $dP/d\varepsilon$ $<$ $dP/d\delta$. In contrast, the $P$ of mutual guiding is higher than that of contact--triggering at a driving force larger than 0.15. Therefore, mutual guiding is superior in ordering efficiency in comparison with contact-triggering when the driving force is sufficiently large. Namely, $dP/d\varepsilon$ $>$ $dP/d\delta$.

To quantitatively confirm this observation of efficiency, we evaluate the efficiencies $dP/d\varepsilon$ and $dP/d\delta$ as rough indicators. We assume that the order parameter $P$ has threshold values, $\varepsilon_c$ and $\delta_c$, and linearly increases above them. On the basis of this assumption, we fit the $P$ as functions of driving forces using the linear function by  the least squares method. For this fitting, we set the range from 0.125 to 0.20 for $\varepsilon$ and $\delta$ to the common onset range of motion ordering and calculate the linear slopes as the efficiencies $dP/d\varepsilon$ and $dP/d\delta$. We obtain $dP/d\varepsilon$ = 8.8 with the threshold $\varepsilon_c$ = 0.12 and $dP/d\delta$ = 4.0 with the threshold $\delta_c$ = 0.11 and show the corresponding lines in Fig.~\ref{fig:comp_suspension}(a). Namely, the efficiency of mutual guiding is superior to that of contact--triggering, $dP/d\varepsilon$ $>$ $dP/d\delta$. The values support the above discussion of comparison of observations. 

Next, we focus on the collective velocity $v$. In Fig.~\ref{fig:comp_suspension}(b), we show $v$ as a function of $\varepsilon$ for mutual guiding and $\delta$ for contact--triggering. $v$ has a threshold value of driving force and then linearly increases with the driving forces $\varepsilon$ and $\delta$. $v$ for mutual guiding is higher than that for contact--triggering at least above the threshold of $\varepsilon$ and $\delta$ $\simeq$ 0.15. Therefore, for the driving force, mutual guiding is more effective than contact--triggering. This quantitative difference is expected to reflect the difference in motion--ordering efficiency.

\begin{figure*}[t]
    \begin{center}
        \includegraphics[width=0.85\linewidth]{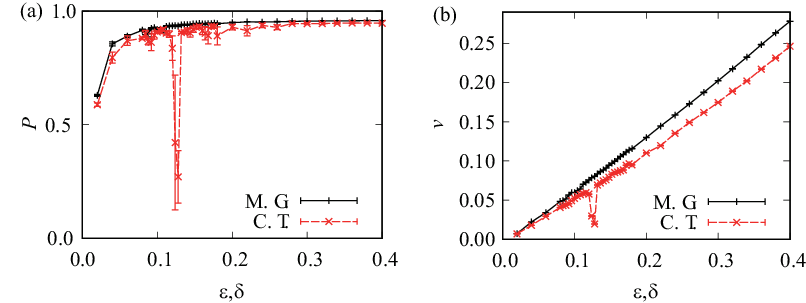}
        \caption{(a) Order parameters $P$ and (b) collective velocity $v$ as a function of $\varepsilon$ for mutual guiding (M. G.) and $\delta$ for contact-triggering (C. T.) for  $\gamma_{\rm E}$ = 4.0 and $N$ = 576 ($\phi$ = 100\%). }\label{fig:comp_aggregation}
    \end{center}
\end{figure*}

Lastly, we examine the cases of large $\gamma_{\rm E}$, where $P$ is the same for both mutual guiding and contact--triggering, as shown in Fig.~\ref{fig:order_parameter}(a). This is also visible in the plot of $P$ in Fig.~\ref{fig:comp_aggregation}(a). However, this does not always indicate that the difference in collective velocity $v$ is also absent in these mechanisms. To examine the difference in $v$, we plot $v$ as a function of the driving force strengths in Fig.~\ref{fig:comp_aggregation}(b). The curves of $v$ seem to have no range of $v$ = 0 and imply a linear response commonly for mutual guiding and contact--triggering. In particular, the value of $v$ is clearly larger than that at a low $\gamma_{\rm E}$. 

A sharp drop in $v$ starts at a certain value of $\delta$ $\simeq$ 0.12 for contact--triggering with $P$ sharply dropping. A similar result is observed in a self-propelled system \citep{Bi:2016,Matsushita:2021a,Pinheiro:2024}. In that self-propelled system, the decrease in $P$ reflects the melting transition from solid to fluid, which implies a similar transition for contact--triggering. This result implies that the cell configuration for contact--triggering is unstable compared with that in mutual guiding.

\section{Summary and Remarks}

We compare mutual guiding with contact--triggering in terms of motion--ordering efficiency. Our main purpose is to confirm the motion--ordering efficiency of cell-to-cell information transfer for mutual guiding. Our results show that the motion--ordering efficiency is almost high for mutual guiding. Exceptionally, mutual guiding is inferior at a low $\gamma_{\rm E}$ in a very narrow region around the onset of driving force. For collective velocity, mutual guiding is superior to contact--triggering. Before this examination, we expected that the difference between these mechanisms with different symmetries would result in a qualitative difference between them. Contrary to this expectation, the difference due to their interaction is only quantitative in terms of the motion--ordering efficiency, and we cannot clearly find the qualitative difference originating from the symmetry difference.

The comparably higher threshold $\varepsilon_{c}$ = 0.12 for mutual guiding, which is shown in Fig.~\ref{fig:comp_suspension}(a), is a characteristic behavior. One possible origin of this behavior is the configuration property of mutual guiding. The interaction through the polarized adhesion in mutual guiding forms a lateral array of the cell configuration \citep{Beug:1973,Muller:1978,Coates:2001,Matsushita:2017}. Therefore, the lateral array formation may inhibit the ordering of the collective motion. Further study of this possibility needs more detailed simulations around the threshold, which remains a future work.

Another peculiar finding is the increase in velocity with $\gamma_{\rm E}$, which is confirmed by the comparison between Figs.~\ref{fig:comp_suspension}(b) and \ref{fig:comp_aggregation}(b). The possible origin of this tendency may be the difference between the suspension and aggregation states of cells. To confirm the effect of the difference between these states, we should examine the changes in $\gamma_{\rm C}$. This is because the dependence on $\gamma_{\rm E}$ is determined by relative magnitude to $\gamma_{\rm C}$. Such a study is also one of our future works.

We consider that the mechanism of contact--triggering may correspond to the contact attraction of locomotion. The contact attraction of locomotion induces a variety of collective motions in synergy with contact following \citep{Hiraiwa:2020}. A previous work showed the absence of collective cell motion without contact following. The previous work focused on the chemotactic response and investigated limited cases where a strong chemotactic response appears. As a result, the collective motion in the case of aggregation or a sufficiently large area fraction was omitted. To confirm this, revising the contact attraction of locomotion in the case of high cell density or aggregation may bring about unknown motion ordering. 

We thank S. Yabunaka, H. Kuwayama, H. Hashimura, M.~Matsumoto, M.~Sawada, and K. Sawamoto for providing much related information. We also thank M. Kikuchi and H. Yoshino for supporting this work. This work was also supported by JSPS KAKENHI (Grant Number 19K03770, 23K03342) and AMED (Grant Number JP19gm1210007).


\begin{thebibliography}{77}%
\makeatletter
\providecommand \@ifxundefined [1]{%
 \@ifx{#1\undefined}
}%
\providecommand \@ifnum [1]{%
 \ifnum #1\expandafter \@firstoftwo
 \else \expandafter \@secondoftwo
 \fi
}%
\providecommand \@ifx [1]{%
 \ifx #1\expandafter \@firstoftwo
 \else \expandafter \@secondoftwo
 \fi
}%
\providecommand \natexlab [1]{#1}%
\providecommand \enquote  [1]{``#1''}%
\providecommand \bibnamefont  [1]{#1}%
\providecommand \bibfnamefont [1]{#1}%
\providecommand \citenamefont [1]{#1}%
\providecommand \href@noop [0]{\@secondoftwo}%
\providecommand \href [0]{\begingroup \@sanitize@url \@href}%
\providecommand \@href[1]{\@@startlink{#1}\@@href}%
\providecommand \@@href[1]{\endgroup#1\@@endlink}%
\providecommand \@sanitize@url [0]{\catcode `\\12\catcode `\$12\catcode
  `\&12\catcode `\#12\catcode `\^12\catcode `\_12\catcode `\%12\relax}%
\providecommand \@@startlink[1]{}%
\providecommand \@@endlink[0]{}%
\providecommand \url  [0]{\begingroup\@sanitize@url \@url }%
\providecommand \@url [1]{\endgroup\@href {#1}{\urlprefix }}%
\providecommand \urlprefix  [0]{URL }%
\providecommand \Eprint [0]{\href }%
\providecommand \doibase [0]{https://doi.org/}%
\providecommand \selectlanguage [0]{\@gobble}%
\providecommand \bibinfo  [0]{\@secondoftwo}%
\providecommand \bibfield  [0]{\@secondoftwo}%
\providecommand \translation [1]{[#1]}%
\providecommand \BibitemOpen [0]{}%
\providecommand \bibitemStop [0]{}%
\providecommand \bibitemNoStop [0]{.\EOS\space}%
\providecommand \EOS [0]{\spacefactor3000\relax}%
\providecommand \BibitemShut  [1]{\csname bibitem#1\endcsname}%
\let\auto@bib@innerbib\@empty
\bibitem [{\citenamefont {R\mbox{\o}rth}(2009)}]{Rorth:2009}%
  \BibitemOpen
  \bibfield  {author} {\bibinfo {author} {\bibfnamefont {P.}~\bibnamefont
  {R\mbox{\o}rth}},\ }\href@noop {} {\bibfield  {journal} {\bibinfo  {journal}
  {Annu. Rev. Cell Dev. Biol.}\ }\textbf {\bibinfo {volume} {25}},\ \bibinfo
  {pages} {407} (\bibinfo {year} {2009})}\BibitemShut {NoStop}%
\bibitem [{\citenamefont {Friedl}\ and\ \citenamefont
  {Gilmour}(2009)}]{Friedl:2009}%
  \BibitemOpen
  \bibfield  {author} {\bibinfo {author} {\bibfnamefont {P.}~\bibnamefont
  {Friedl}}\ and\ \bibinfo {author} {\bibfnamefont {D.}~\bibnamefont
  {Gilmour}},\ }\href@noop {} {\bibfield  {journal} {\bibinfo  {journal} {Nat.
  Rev. Mol. Cell Biol.}\ }\textbf {\bibinfo {volume} {10}},\ \bibinfo {pages}
  {445} (\bibinfo {year} {2009})}\BibitemShut {NoStop}%
\bibitem [{\citenamefont {Weijer}(2009)}]{Weijer:2009}%
  \BibitemOpen
  \bibfield  {author} {\bibinfo {author} {\bibfnamefont {C.~J.}\ \bibnamefont
  {Weijer}},\ }\href@noop {} {\bibfield  {journal} {\bibinfo  {journal} {J.
  Cell Sci.}\ }\textbf {\bibinfo {volume} {122}},\ \bibinfo {pages} {3215}
  (\bibinfo {year} {2009})}\BibitemShut {NoStop}%
\bibitem [{\citenamefont {Haeger}\ \emph {et~al.}(2015)\citenamefont {Haeger},
  \citenamefont {Wolf}, \citenamefont {Zegers},\ and\ \citenamefont
  {Friedl}}]{Haeger:2015}%
  \BibitemOpen
  \bibfield  {author} {\bibinfo {author} {\bibfnamefont {A.}~\bibnamefont
  {Haeger}}, \bibinfo {author} {\bibfnamefont {K.}~\bibnamefont {Wolf}},
  \bibinfo {author} {\bibfnamefont {M.~M.}\ \bibnamefont {Zegers}},\ and\
  \bibinfo {author} {\bibfnamefont {P.}~\bibnamefont {Friedl}},\ }\href@noop {}
  {\bibfield  {journal} {\bibinfo  {journal} {Trends Cell Biol.}\ }\textbf
  {\bibinfo {volume} {25}},\ \bibinfo {pages} {556} (\bibinfo {year}
  {2015})}\BibitemShut {NoStop}%
\bibitem [{\citenamefont {Coates}\ and\ \citenamefont
  {Harwood}(2001)}]{Coates:2001}%
  \BibitemOpen
  \bibfield  {author} {\bibinfo {author} {\bibfnamefont {J.~C.}\ \bibnamefont
  {Coates}}\ and\ \bibinfo {author} {\bibfnamefont {A.~J.}\ \bibnamefont
  {Harwood}},\ }\href@noop {} {\bibfield  {journal} {\bibinfo  {journal} {J.
  Cell Sci.}\ }\textbf {\bibinfo {volume} {114}},\ \bibinfo {pages} {4349}
  (\bibinfo {year} {2001})}\BibitemShut {NoStop}%
\bibitem [{\citenamefont {Stramer}\ and\ \citenamefont
  {Mayor}(2016)}]{Stramer:2016}%
  \BibitemOpen
  \bibfield  {author} {\bibinfo {author} {\bibfnamefont {B.}~\bibnamefont
  {Stramer}}\ and\ \bibinfo {author} {\bibfnamefont {R.}~\bibnamefont
  {Mayor}},\ }\href@noop {} {\bibfield  {journal} {\bibinfo  {journal} {Nat.
  Rev. Mol. Cell Biol.}\ }\textbf {\bibinfo {volume} {18}},\ \bibinfo {pages}
  {43} (\bibinfo {year} {2016})}\BibitemShut {NoStop}%
\bibitem [{\citenamefont {Marchetti}\ \emph {et~al.}(2013)\citenamefont
  {Marchetti}, \citenamefont {Joanny}, \citenamefont {Ramaswamy}, \citenamefont
  {Liverpool}, \citenamefont {Prost}, \citenamefont {Rao},\ and\ \citenamefont
  {Simha}}]{Marchetti:2013}%
  \BibitemOpen
  \bibfield  {author} {\bibinfo {author} {\bibfnamefont {M.~C.}\ \bibnamefont
  {Marchetti}}, \bibinfo {author} {\bibfnamefont {J.~F.}\ \bibnamefont
  {Joanny}}, \bibinfo {author} {\bibfnamefont {S.}~\bibnamefont {Ramaswamy}},
  \bibinfo {author} {\bibfnamefont {T.~B.}\ \bibnamefont {Liverpool}}, \bibinfo
  {author} {\bibfnamefont {J.}~\bibnamefont {Prost}}, \bibinfo {author}
  {\bibfnamefont {M.}~\bibnamefont {Rao}},\ and\ \bibinfo {author}
  {\bibfnamefont {R.~A.}\ \bibnamefont {Simha}},\ }\href@noop {} {\bibfield
  {journal} {\bibinfo  {journal} {Rev. Mod. Phys.}\ }\textbf {\bibinfo {volume}
  {85}},\ \bibinfo {pages} {1143} (\bibinfo {year} {2013})}\BibitemShut
  {NoStop}%
\bibitem [{\citenamefont {Hakim}\ and\ \citenamefont
  {Silberzan}(2017)}]{Hakim:2017}%
  \BibitemOpen
  \bibfield  {author} {\bibinfo {author} {\bibfnamefont {V.}~\bibnamefont
  {Hakim}}\ and\ \bibinfo {author} {\bibfnamefont {P.}~\bibnamefont
  {Silberzan}},\ }\href@noop {} {\bibfield  {journal} {\bibinfo  {journal}
  {Rep. Prog. Phys.}\ }\textbf {\bibinfo {volume} {80}},\ \bibinfo {pages}
  {076601} (\bibinfo {year} {2017})}\BibitemShut {NoStop}%
\bibitem [{\citenamefont {Swaney}\ \emph {et~al.}(2010)\citenamefont {Swaney},
  \citenamefont {Huang},\ and\ \citenamefont {Devreotes}}]{Swaney:2010}%
  \BibitemOpen
  \bibfield  {author} {\bibinfo {author} {\bibfnamefont {K.~F.}\ \bibnamefont
  {Swaney}}, \bibinfo {author} {\bibfnamefont {C.-H.}\ \bibnamefont {Huang}},\
  and\ \bibinfo {author} {\bibfnamefont {P.~N.}\ \bibnamefont {Devreotes}},\
  }\href@noop {} {\bibfield  {journal} {\bibinfo  {journal} {Annu. Rev.
  Biophys.}\ }\textbf {\bibinfo {volume} {39}},\ \bibinfo {pages} {265}
  (\bibinfo {year} {2010})}\BibitemShut {NoStop}%
\bibitem [{\citenamefont {Camley}\ \emph {et~al.}(2016)\citenamefont {Camley},
  \citenamefont {Zimmermann}, \citenamefont {Levine},\ and\ \citenamefont
  {Rappel}}]{Camley:2016}%
  \BibitemOpen
  \bibfield  {author} {\bibinfo {author} {\bibfnamefont {B.~A.}\ \bibnamefont
  {Camley}}, \bibinfo {author} {\bibfnamefont {J.}~\bibnamefont {Zimmermann}},
  \bibinfo {author} {\bibfnamefont {H.}~\bibnamefont {Levine}},\ and\ \bibinfo
  {author} {\bibfnamefont {W.-J.}\ \bibnamefont {Rappel}},\ }\href@noop {}
  {\bibfield  {journal} {\bibinfo  {journal} {Phys. Rev. Lett.}\ }\textbf
  {\bibinfo {volume} {116}},\ \bibinfo {pages} {098101} (\bibinfo {year}
  {2016})}\BibitemShut {NoStop}%
\bibitem [{\citenamefont {Camley}(2018)}]{Camley:2018}%
  \BibitemOpen
  \bibfield  {author} {\bibinfo {author} {\bibfnamefont {B.~A.}\ \bibnamefont
  {Camley}},\ }\href@noop {} {\bibfield  {journal} {\bibinfo  {journal} {J.
  Phys. Condens. Matter}\ }\textbf {\bibinfo {volume} {30}},\ \bibinfo {pages}
  {223001.} (\bibinfo {year} {2018})}\BibitemShut {NoStop}%
\bibitem [{\citenamefont {Fujimori}\ \emph {et~al.}(2019)\citenamefont
  {Fujimori}, \citenamefont {Nakajima}, \citenamefont {Shimada},\ and\
  \citenamefont {Sawai}}]{Fujimori:2019}%
  \BibitemOpen
  \bibfield  {author} {\bibinfo {author} {\bibfnamefont {T.}~\bibnamefont
  {Fujimori}}, \bibinfo {author} {\bibfnamefont {A.}~\bibnamefont {Nakajima}},
  \bibinfo {author} {\bibfnamefont {N.}~\bibnamefont {Shimada}},\ and\ \bibinfo
  {author} {\bibfnamefont {S.}~\bibnamefont {Sawai}},\ }\href@noop {}
  {\bibfield  {journal} {\bibinfo  {journal} {Proc. Natl. Acad. Sci. USA}\
  }\textbf {\bibinfo {volume} {116}},\ \bibinfo {pages} {4291} (\bibinfo {year}
  {2019})}\BibitemShut {NoStop}%
\bibitem [{\citenamefont {Hiraiwa}(2020)}]{Hiraiwa:2020}%
  \BibitemOpen
  \bibfield  {author} {\bibinfo {author} {\bibfnamefont {T.}~\bibnamefont
  {Hiraiwa}},\ }\href@noop {} {\bibfield  {journal} {\bibinfo  {journal} {Phys.
  Rev. Lett.}\ }\textbf {\bibinfo {volume} {125}},\ \bibinfo {pages} {268104}
  (\bibinfo {year} {2020})}\BibitemShut {NoStop}%
\bibitem [{\citenamefont {Hayakawa}\ \emph {et~al.}(2020)\citenamefont
  {Hayakawa}, \citenamefont {Hiraiwa}, \citenamefont {Wada}, \citenamefont
  {Kuwayama},\ and\ \citenamefont {Shibata}}]{Hayakawa:2020}%
  \BibitemOpen
  \bibfield  {author} {\bibinfo {author} {\bibfnamefont {M.}~\bibnamefont
  {Hayakawa}}, \bibinfo {author} {\bibfnamefont {T.}~\bibnamefont {Hiraiwa}},
  \bibinfo {author} {\bibfnamefont {Y.}~\bibnamefont {Wada}}, \bibinfo {author}
  {\bibfnamefont {H.}~\bibnamefont {Kuwayama}},\ and\ \bibinfo {author}
  {\bibfnamefont {T.}~\bibnamefont {Shibata}},\ }\href@noop {} {\bibfield
  {journal} {\bibinfo  {journal} {eLife}\ }\textbf {\bibinfo {volume} {9}},\
  \bibinfo {pages} {e53609} (\bibinfo {year} {2020})}\BibitemShut {NoStop}%
\bibitem [{\citenamefont {Schnyder}\ \emph {et~al.}(2017)\citenamefont
  {Schnyder}, \citenamefont {Molina}, \citenamefont {Tanaka},\ and\
  \citenamefont {Yamamoto}}]{Schnyder:2017}%
  \BibitemOpen
  \bibfield  {author} {\bibinfo {author} {\bibfnamefont {S.~K.}\ \bibnamefont
  {Schnyder}}, \bibinfo {author} {\bibfnamefont {J.~J.}\ \bibnamefont
  {Molina}}, \bibinfo {author} {\bibfnamefont {Y.}~\bibnamefont {Tanaka}},\
  and\ \bibinfo {author} {\bibfnamefont {R.}~\bibnamefont {Yamamoto}},\
  }\href@noop {} {\bibfield  {journal} {\bibinfo  {journal} {Sci. Rep.}\
  }\textbf {\bibinfo {volume} {7}},\ \bibinfo {pages} {5163} (\bibinfo {year}
  {2017})}\BibitemShut {NoStop}%
\bibitem [{\citenamefont {Hiraiwa}(2019)}]{Hiraiwa:2019}%
  \BibitemOpen
  \bibfield  {author} {\bibinfo {author} {\bibfnamefont {T.}~\bibnamefont
  {Hiraiwa}},\ }\href@noop {} {\bibfield  {journal} {\bibinfo  {journal} {Phys.
  Rev. E}\ }\textbf {\bibinfo {volume} {99}},\ \bibinfo {pages} {012614}
  (\bibinfo {year} {2019})}\BibitemShut {NoStop}%
\bibitem [{\citenamefont {Jain}\ \emph {et~al.}(2022)\citenamefont {Jain},
  \citenamefont {Wenzel},\ and\ \citenamefont {Voigt}}]{Jain:2022}%
  \BibitemOpen
  \bibfield  {author} {\bibinfo {author} {\bibfnamefont {H.~P.}\ \bibnamefont
  {Jain}}, \bibinfo {author} {\bibfnamefont {D.}~\bibnamefont {Wenzel}},\ and\
  \bibinfo {author} {\bibfnamefont {A.}~\bibnamefont {Voigt}},\ }\href@noop {}
  {\bibfield  {journal} {\bibinfo  {journal} {Phys. Rev. E}\ }\textbf {\bibinfo
  {volume} {105}},\ \bibinfo {pages} {034402} (\bibinfo {year}
  {2022})}\BibitemShut {NoStop}%
\bibitem [{\citenamefont {Omelchenko}\ \emph {et~al.}(2003)\citenamefont
  {Omelchenko}, \citenamefont {Vasiliev}, \citenamefont {Gelfand},
  \citenamefont {Feder},\ and\ \citenamefont {Bonder}}]{Omelchenko:2003}%
  \BibitemOpen
  \bibfield  {author} {\bibinfo {author} {\bibfnamefont {T.}~\bibnamefont
  {Omelchenko}}, \bibinfo {author} {\bibfnamefont {J.~M.}\ \bibnamefont
  {Vasiliev}}, \bibinfo {author} {\bibfnamefont {I.~M.}\ \bibnamefont
  {Gelfand}}, \bibinfo {author} {\bibfnamefont {H.~H.}\ \bibnamefont {Feder}},\
  and\ \bibinfo {author} {\bibfnamefont {E.~M.}\ \bibnamefont {Bonder}},\
  }\href@noop {} {\bibfield  {journal} {\bibinfo  {journal} {Proc. Natl Acad.
  Sci. USA}\ }\textbf {\bibinfo {volume} {100}},\ \bibinfo {pages} {10788}
  (\bibinfo {year} {2003})}\BibitemShut {NoStop}%
\bibitem [{\citenamefont {Haga}\ \emph {et~al.}(2005)\citenamefont {Haga},
  \citenamefont {Irahara}, \citenamefont {Kobayashi}, \citenamefont
  {Nakagaki},\ and\ \citenamefont {Kawabata}}]{Haga:2005}%
  \BibitemOpen
  \bibfield  {author} {\bibinfo {author} {\bibfnamefont {H.}~\bibnamefont
  {Haga}}, \bibinfo {author} {\bibfnamefont {C.}~\bibnamefont {Irahara}},
  \bibinfo {author} {\bibfnamefont {R.}~\bibnamefont {Kobayashi}}, \bibinfo
  {author} {\bibfnamefont {T.}~\bibnamefont {Nakagaki}},\ and\ \bibinfo
  {author} {\bibfnamefont {K.}~\bibnamefont {Kawabata}},\ }\href@noop {}
  {\bibfield  {journal} {\bibinfo  {journal} {biophys. J.}\ }\textbf {\bibinfo
  {volume} {88}},\ \bibinfo {pages} {2250} (\bibinfo {year}
  {2005})}\BibitemShut {NoStop}%
\bibitem [{\citenamefont {Trepat}\ \emph {et~al.}(2009)\citenamefont {Trepat},
  \citenamefont {Wasserman}, \citenamefont {Angelini}, \citenamefont {Millet},
  \citenamefont {Weitz}, \citenamefont {Butler},\ and\ \citenamefont
  {Fredberg}}]{Trepat:2009}%
  \BibitemOpen
  \bibfield  {author} {\bibinfo {author} {\bibfnamefont {X.}~\bibnamefont
  {Trepat}}, \bibinfo {author} {\bibfnamefont {M.~R.}\ \bibnamefont
  {Wasserman}}, \bibinfo {author} {\bibfnamefont {T.~E.}\ \bibnamefont
  {Angelini}}, \bibinfo {author} {\bibfnamefont {E.}~\bibnamefont {Millet}},
  \bibinfo {author} {\bibfnamefont {D.~A.}\ \bibnamefont {Weitz}}, \bibinfo
  {author} {\bibfnamefont {J.~P.}\ \bibnamefont {Butler}},\ and\ \bibinfo
  {author} {\bibfnamefont {J.~J.}\ \bibnamefont {Fredberg}},\ }\href@noop {}
  {\bibfield  {journal} {\bibinfo  {journal} {Nat. Phys.}\ }\textbf {\bibinfo
  {volume} {5}},\ \bibinfo {pages} {426} (\bibinfo {year} {2009})}\BibitemShut
  {NoStop}%
\bibitem [{\citenamefont {Kabla}(2012)}]{Kabla:2012}%
  \BibitemOpen
  \bibfield  {author} {\bibinfo {author} {\bibfnamefont {A.~J.}\ \bibnamefont
  {Kabla}},\ }\href@noop {} {\bibfield  {journal} {\bibinfo  {journal} {J. R.
  Soc. Interface}\ }\textbf {\bibinfo {volume} {9}},\ \bibinfo {pages} {3268}
  (\bibinfo {year} {2012})}\BibitemShut {NoStop}%
\bibitem [{\citenamefont {Serra-Picamal}\ \emph {et~al.}(2012)\citenamefont
  {Serra-Picamal}, \citenamefont {Conte}, \citenamefont {Vincent},
  \citenamefont {Anon}, \citenamefont {Tambe}, \citenamefont {Bazellieres},
  \citenamefont {Butler}, \citenamefont {Fredberg},\ and\ \citenamefont
  {Trepat}}]{Serra-Picamal:2012}%
  \BibitemOpen
  \bibfield  {author} {\bibinfo {author} {\bibfnamefont {X.}~\bibnamefont
  {Serra-Picamal}}, \bibinfo {author} {\bibfnamefont {V.}~\bibnamefont
  {Conte}}, \bibinfo {author} {\bibfnamefont {R.}~\bibnamefont {Vincent}},
  \bibinfo {author} {\bibfnamefont {E.}~\bibnamefont {Anon}}, \bibinfo {author}
  {\bibfnamefont {D.~T.}\ \bibnamefont {Tambe}}, \bibinfo {author}
  {\bibfnamefont {E.}~\bibnamefont {Bazellieres}}, \bibinfo {author}
  {\bibfnamefont {J.~P.}\ \bibnamefont {Butler}}, \bibinfo {author}
  {\bibfnamefont {J.~J.}\ \bibnamefont {Fredberg}},\ and\ \bibinfo {author}
  {\bibfnamefont {X.}~\bibnamefont {Trepat}},\ }\href@noop {} {\bibfield
  {journal} {\bibinfo  {journal} {Nat. Phys.}\ }\textbf {\bibinfo {volume}
  {8}},\ \bibinfo {pages} {628} (\bibinfo {year} {2012})}\BibitemShut {NoStop}%
\bibitem [{\citenamefont {Notbohm}\ \emph {et~al.}(2016)\citenamefont
  {Notbohm}, \citenamefont {Banerjee}, \citenamefont {Utuje}, \citenamefont
  {Gweon}, \citenamefont {Jang}, \citenamefont {Park}, \citenamefont {Shin},
  \citenamefont {Butler}, \citenamefont {Fredberg},\ and\ \citenamefont
  {Marchetti}}]{Notbohm:2016}%
  \BibitemOpen
  \bibfield  {author} {\bibinfo {author} {\bibfnamefont {J.}~\bibnamefont
  {Notbohm}}, \bibinfo {author} {\bibfnamefont {S.}~\bibnamefont {Banerjee}},
  \bibinfo {author} {\bibfnamefont {K.~J.~C.}\ \bibnamefont {Utuje}}, \bibinfo
  {author} {\bibfnamefont {B.}~\bibnamefont {Gweon}}, \bibinfo {author}
  {\bibfnamefont {H.}~\bibnamefont {Jang}}, \bibinfo {author} {\bibfnamefont
  {Y.}~\bibnamefont {Park}}, \bibinfo {author} {\bibfnamefont {J.}~\bibnamefont
  {Shin}}, \bibinfo {author} {\bibfnamefont {J.~P.}\ \bibnamefont {Butler}},
  \bibinfo {author} {\bibfnamefont {J.~J.}\ \bibnamefont {Fredberg}},\ and\
  \bibinfo {author} {\bibfnamefont {M.~C.}\ \bibnamefont {Marchetti}},\
  }\href@noop {} {\bibfield  {journal} {\bibinfo  {journal} {Biophys. J.}\
  }\textbf {\bibinfo {volume} {110}},\ \bibinfo {pages} {2729} (\bibinfo {year}
  {2016})}\BibitemShut {NoStop}%
\bibitem [{\citenamefont {Yabunaka}\ and\ \citenamefont
  {Marcq}(2017{\natexlab{a}})}]{Yabunaka:2017a}%
  \BibitemOpen
  \bibfield  {author} {\bibinfo {author} {\bibfnamefont {S.}~\bibnamefont
  {Yabunaka}}\ and\ \bibinfo {author} {\bibfnamefont {P.}~\bibnamefont
  {Marcq}},\ }\href@noop {} {\bibfield  {journal} {\bibinfo  {journal} {Phys.
  Rev. E}\ }\textbf {\bibinfo {volume} {96}},\ \bibinfo {pages} {022406}
  (\bibinfo {year} {2017}{\natexlab{a}})}\BibitemShut {NoStop}%
\bibitem [{\citenamefont {Yabunaka}\ and\ \citenamefont
  {Marcq}(2017{\natexlab{b}})}]{Yabunaka:2017b}%
  \BibitemOpen
  \bibfield  {author} {\bibinfo {author} {\bibfnamefont {S.}~\bibnamefont
  {Yabunaka}}\ and\ \bibinfo {author} {\bibfnamefont {P.}~\bibnamefont
  {Marcq}},\ }\href@noop {} {\bibfield  {journal} {\bibinfo  {journal} {Soft
  Matter}\ }\textbf {\bibinfo {volume} {13}},\ \bibinfo {pages} {7046}
  (\bibinfo {year} {2017}{\natexlab{b}})}\BibitemShut {NoStop}%
\bibitem [{\citenamefont {Tlili}\ \emph {et~al.}(2018)\citenamefont {Tlili},
  \citenamefont {Gauquelin}, \citenamefont {Li}, \citenamefont {Cardoso},
  \citenamefont {Beno\^{i}t~Ladoux},\ and\ \citenamefont
  {Graner}}]{Tlili:2018}%
  \BibitemOpen
  \bibfield  {author} {\bibinfo {author} {\bibfnamefont {S.}~\bibnamefont
  {Tlili}}, \bibinfo {author} {\bibfnamefont {E.}~\bibnamefont {Gauquelin}},
  \bibinfo {author} {\bibfnamefont {B.}~\bibnamefont {Li}}, \bibinfo {author}
  {\bibfnamefont {O.}~\bibnamefont {Cardoso}}, \bibinfo {author} {\bibfnamefont
  {H.~D.-A.}\ \bibnamefont {Beno\^{i}t~Ladoux}},\ and\ \bibinfo {author}
  {\bibfnamefont {F.}~\bibnamefont {Graner}},\ }\href@noop {} {\bibfield
  {journal} {\bibinfo  {journal} {R. Soc. Op. Sci.}\ }\textbf {\bibinfo
  {volume} {5}},\ \bibinfo {pages} {172421} (\bibinfo {year}
  {2018})}\BibitemShut {NoStop}%
\bibitem [{\citenamefont {Fukuyama}\ \emph {et~al.}(2020)\citenamefont
  {Fukuyama}, \citenamefont {Ebata}, \citenamefont {Kondo}, \citenamefont
  {Kidoaki}, \citenamefont {Aoki},\ and\ \citenamefont
  {Maeda}}]{Fukuyama:2020}%
  \BibitemOpen
  \bibfield  {author} {\bibinfo {author} {\bibfnamefont {T.}~\bibnamefont
  {Fukuyama}}, \bibinfo {author} {\bibfnamefont {H.}~\bibnamefont {Ebata}},
  \bibinfo {author} {\bibfnamefont {Y.}~\bibnamefont {Kondo}}, \bibinfo
  {author} {\bibfnamefont {S.}~\bibnamefont {Kidoaki}}, \bibinfo {author}
  {\bibfnamefont {K.}~\bibnamefont {Aoki}},\ and\ \bibinfo {author}
  {\bibfnamefont {Y.~T.}\ \bibnamefont {Maeda}}} (\bibinfo {year} {2020}),\
  \bibinfo {note} {arXiv:2008.12955}\BibitemShut {NoStop}%
\bibitem [{\citenamefont {Li}\ \emph {et~al.}(2021)\citenamefont {Li},
  \citenamefont {Schnyder}, \citenamefont {Turner},\ and\ \citenamefont
  {Yamamoto}}]{Li:2021}%
  \BibitemOpen
  \bibfield  {author} {\bibinfo {author} {\bibfnamefont {J.}~\bibnamefont
  {Li}}, \bibinfo {author} {\bibfnamefont {S.~K.}\ \bibnamefont {Schnyder}},
  \bibinfo {author} {\bibfnamefont {M.~S.}\ \bibnamefont {Turner}},\ and\
  \bibinfo {author} {\bibfnamefont {R.}~\bibnamefont {Yamamoto}},\ }\href@noop
  {} {\bibfield  {journal} {\bibinfo  {journal} {Phys. Rev. X}\ }\textbf
  {\bibinfo {volume} {11}},\ \bibinfo {pages} {031025} (\bibinfo {year}
  {2021})}\BibitemShut {NoStop}%
\bibitem [{\citenamefont {Alhashem}\ \emph {et~al.}(2022)\citenamefont
  {Alhashem}, \citenamefont {Feldner-Busztin}, \citenamefont {Revell},
  \citenamefont {Portillo}, \citenamefont {Camargo-Sosa}, \citenamefont
  {Richardson}, \citenamefont {Rocha}, \citenamefont {Gauert}, \citenamefont
  {Corbeaux},\ and\ \citenamefont {Linker}}]{Alhashem:2022}%
  \BibitemOpen
  \bibfield  {author} {\bibinfo {author} {\bibfnamefont {Z.}~\bibnamefont
  {Alhashem}}, \bibinfo {author} {\bibfnamefont {D.}~\bibnamefont
  {Feldner-Busztin}}, \bibinfo {author} {\bibfnamefont {C.}~\bibnamefont
  {Revell}}, \bibinfo {author} {\bibfnamefont {M.~A.-G.}\ \bibnamefont
  {Portillo}}, \bibinfo {author} {\bibfnamefont {K.}~\bibnamefont
  {Camargo-Sosa}}, \bibinfo {author} {\bibfnamefont {J.}~\bibnamefont
  {Richardson}}, \bibinfo {author} {\bibfnamefont {M.}~\bibnamefont {Rocha}},
  \bibinfo {author} {\bibfnamefont {A.}~\bibnamefont {Gauert}}, \bibinfo
  {author} {\bibfnamefont {T.}~\bibnamefont {Corbeaux}},\ and\ \bibinfo
  {author} {\bibfnamefont {C.}~\bibnamefont {Linker}},\ }\href@noop {}
  {\bibfield  {journal} {\bibinfo  {journal} {eLife}\ }\textbf {\bibinfo
  {volume} {11}},\ \bibinfo {pages} {e73550} (\bibinfo {year}
  {2022})}\BibitemShut {NoStop}%
\bibitem [{\citenamefont {Szab\'{o}}\ \emph {et~al.}(2006)\citenamefont
  {Szab\'{o}}, \citenamefont {Szollosi}, \citenamefont {Gonci}, \citenamefont
  {Juranyi}, \citenamefont {Selmeczi},\ and\ \citenamefont
  {Vicsek}}]{Szabo:2006}%
  \BibitemOpen
  \bibfield  {author} {\bibinfo {author} {\bibfnamefont {B.}~\bibnamefont
  {Szab\'{o}}}, \bibinfo {author} {\bibfnamefont {G.~J.}\ \bibnamefont
  {Szollosi}}, \bibinfo {author} {\bibfnamefont {B.}~\bibnamefont {Gonci}},
  \bibinfo {author} {\bibfnamefont {Z.}~\bibnamefont {Juranyi}}, \bibinfo
  {author} {\bibfnamefont {D.}~\bibnamefont {Selmeczi}},\ and\ \bibinfo
  {author} {\bibfnamefont {T.}~\bibnamefont {Vicsek}},\ }\href@noop {}
  {\bibfield  {journal} {\bibinfo  {journal} {Phys.~Rev.~E}\ }\textbf {\bibinfo
  {volume} {74}},\ \bibinfo {pages} {061908} (\bibinfo {year}
  {2006})}\BibitemShut {NoStop}%
\bibitem [{\citenamefont {Matsushita}\ \emph {et~al.}(2019)\citenamefont
  {Matsushita}, \citenamefont {Horibe}, \citenamefont {Kamamoto},\ and\
  \citenamefont {Fujimoto}}]{Matsushita:2019}%
  \BibitemOpen
  \bibfield  {author} {\bibinfo {author} {\bibfnamefont {K.}~\bibnamefont
  {Matsushita}}, \bibinfo {author} {\bibfnamefont {K.}~\bibnamefont {Horibe}},
  \bibinfo {author} {\bibfnamefont {N.}~\bibnamefont {Kamamoto}},\ and\
  \bibinfo {author} {\bibfnamefont {K.}~\bibnamefont {Fujimoto}},\ }\href@noop
  {} {\bibfield  {journal} {\bibinfo  {journal} {J. Phys. Soc. Jpn.}\ }\textbf
  {\bibinfo {volume} {88}},\ \bibinfo {pages} {103801} (\bibinfo {year}
  {2019})}\BibitemShut {NoStop}%
\bibitem [{\citenamefont {Matsushita}(2020)}]{Matsushita:2020a}%
  \BibitemOpen
  \bibfield  {author} {\bibinfo {author} {\bibfnamefont {K.}~\bibnamefont
  {Matsushita}},\ }\href@noop {} {\bibfield  {journal} {\bibinfo  {journal}
  {Phys. Rev. E}\ }\textbf {\bibinfo {volume} {101}},\ \bibinfo {pages}
  {052410} (\bibinfo {year} {2020})}\BibitemShut {NoStop}%
\bibitem [{\citenamefont {Sato}\ \emph
  {et~al.}(2015{\natexlab{a}})\citenamefont {Sato}, \citenamefont {Hiraiwa},\
  and\ \citenamefont {Shibata}}]{Sato:2015a}%
  \BibitemOpen
  \bibfield  {author} {\bibinfo {author} {\bibfnamefont {K.}~\bibnamefont
  {Sato}}, \bibinfo {author} {\bibfnamefont {T.}~\bibnamefont {Hiraiwa}},\ and\
  \bibinfo {author} {\bibfnamefont {T.}~\bibnamefont {Shibata}},\ }\href@noop
  {} {\bibfield  {journal} {\bibinfo  {journal} {Phys. Rev. Lett.}\ }\textbf
  {\bibinfo {volume} {115}},\ \bibinfo {pages} {188102} (\bibinfo {year}
  {2015}{\natexlab{a}})}\BibitemShut {NoStop}%
\bibitem [{\citenamefont {Sato}\ \emph
  {et~al.}(2015{\natexlab{b}})\citenamefont {Sato}, \citenamefont {Hiraiwa},
  \citenamefont {Maekawa}, \citenamefont {Isomura}, \citenamefont {Shibata},\
  and\ \citenamefont {Kuranaga}}]{Sato:2015b}%
  \BibitemOpen
  \bibfield  {author} {\bibinfo {author} {\bibfnamefont {K.}~\bibnamefont
  {Sato}}, \bibinfo {author} {\bibfnamefont {T.}~\bibnamefont {Hiraiwa}},
  \bibinfo {author} {\bibfnamefont {E.}~\bibnamefont {Maekawa}}, \bibinfo
  {author} {\bibfnamefont {A.}~\bibnamefont {Isomura}}, \bibinfo {author}
  {\bibfnamefont {T.}~\bibnamefont {Shibata}},\ and\ \bibinfo {author}
  {\bibfnamefont {E.}~\bibnamefont {Kuranaga}},\ }\href@noop {} {\bibfield
  {journal} {\bibinfo  {journal} {Nat. Commun.}\ }\textbf {\bibinfo {volume}
  {6}},\ \bibinfo {pages} {10074} (\bibinfo {year}
  {2015}{\natexlab{b}})}\BibitemShut {NoStop}%
\bibitem [{\citenamefont {Lo}\ \emph {et~al.}(2000)\citenamefont {Lo},
  \citenamefont {Wang}, \citenamefont {Dembo},\ and\ \citenamefont
  {Wang}}]{Lo:2000}%
  \BibitemOpen
  \bibfield  {author} {\bibinfo {author} {\bibfnamefont {C.-M.}\ \bibnamefont
  {Lo}}, \bibinfo {author} {\bibfnamefont {H.-B.}\ \bibnamefont {Wang}},
  \bibinfo {author} {\bibfnamefont {M.}~\bibnamefont {Dembo}},\ and\ \bibinfo
  {author} {\bibfnamefont {Y.-L.}\ \bibnamefont {Wang}},\ }\href@noop {}
  {\bibfield  {journal} {\bibinfo  {journal} {Biophys. J.}\ }\textbf {\bibinfo
  {volume} {79}},\ \bibinfo {pages} {144} (\bibinfo {year} {2000})}\BibitemShut
  {NoStop}%
\bibitem [{\citenamefont {Sunyer}\ \emph {et~al.}(2016)\citenamefont {Sunyer},
  \citenamefont {Conte}, \citenamefont {Escribano}, \citenamefont
  {Elosegui-Artola}, \citenamefont {Labernadie}, \citenamefont {Valon},
  \citenamefont {Navajas}, \citenamefont {Garc\'{i}a-Aznar}, \citenamefont
  {Mu{\~n}oz}, \citenamefont {Roca-Cusachs},\ and\ \citenamefont
  {Trepat}}]{Sunyer:2016}%
  \BibitemOpen
  \bibfield  {author} {\bibinfo {author} {\bibfnamefont {R.}~\bibnamefont
  {Sunyer}}, \bibinfo {author} {\bibfnamefont {V.}~\bibnamefont {Conte}},
  \bibinfo {author} {\bibfnamefont {J.}~\bibnamefont {Escribano}}, \bibinfo
  {author} {\bibfnamefont {A.}~\bibnamefont {Elosegui-Artola}}, \bibinfo
  {author} {\bibfnamefont {A.}~\bibnamefont {Labernadie}}, \bibinfo {author}
  {\bibfnamefont {L.}~\bibnamefont {Valon}}, \bibinfo {author} {\bibfnamefont
  {D.}~\bibnamefont {Navajas}}, \bibinfo {author} {\bibfnamefont {J.~M.}\
  \bibnamefont {Garc\'{i}a-Aznar}}, \bibinfo {author} {\bibfnamefont {J.~J.}\
  \bibnamefont {Mu{\~n}oz}}, \bibinfo {author} {\bibfnamefont {P.}~\bibnamefont
  {Roca-Cusachs}},\ and\ \bibinfo {author} {\bibfnamefont {X.}~\bibnamefont
  {Trepat}},\ }\href@noop {} {\bibfield  {journal} {\bibinfo  {journal}
  {Science}\ }\textbf {\bibinfo {volume} {353}},\ \bibinfo {pages} {1157}
  (\bibinfo {year} {2016})}\BibitemShut {NoStop}%
\bibitem [{\citenamefont {Pallar\`{e}s}\ \emph {et~al.}(2023)\citenamefont
  {Pallar\`{e}s}, \citenamefont {Pi-Jaum\`{a}}, \citenamefont {Fortunato},
  \citenamefont {Grazu}, \citenamefont {G\'{o}mez-Gonz\'{o}lez}, \citenamefont
  {Roca-Cusachs}, \citenamefont {de~la Fuente}, \citenamefont {Alert},
  \citenamefont {Sunyer}, \citenamefont {Casademunt},\ and\ \citenamefont
  {Trepat}}]{Pallares:2023}%
  \BibitemOpen
  \bibfield  {author} {\bibinfo {author} {\bibfnamefont {M.~E.}\ \bibnamefont
  {Pallar\`{e}s}}, \bibinfo {author} {\bibfnamefont {I.}~\bibnamefont
  {Pi-Jaum\`{a}}}, \bibinfo {author} {\bibfnamefont {I.~C.}\ \bibnamefont
  {Fortunato}}, \bibinfo {author} {\bibfnamefont {V.}~\bibnamefont {Grazu}},
  \bibinfo {author} {\bibfnamefont {M.}~\bibnamefont {G\'{o}mez-Gonz\'{o}lez}},
  \bibinfo {author} {\bibfnamefont {P.}~\bibnamefont {Roca-Cusachs}}, \bibinfo
  {author} {\bibfnamefont {J.~M.}\ \bibnamefont {de~la Fuente}}, \bibinfo
  {author} {\bibfnamefont {R.}~\bibnamefont {Alert}}, \bibinfo {author}
  {\bibfnamefont {R.}~\bibnamefont {Sunyer}}, \bibinfo {author} {\bibfnamefont
  {J.}~\bibnamefont {Casademunt}},\ and\ \bibinfo {author} {\bibfnamefont
  {X.}~\bibnamefont {Trepat}},\ }\href@noop {} {\bibfield  {journal} {\bibinfo
  {journal} {Nat. Phys.}\ }\textbf {\bibinfo {volume} {19}},\ \bibinfo {pages}
  {279} (\bibinfo {year} {2023})}\BibitemShut {NoStop}%
\bibitem [{\citenamefont {Matsushita}(2018)}]{Matsushita:2018}%
  \BibitemOpen
  \bibfield  {author} {\bibinfo {author} {\bibfnamefont {K.}~\bibnamefont
  {Matsushita}},\ }\href@noop {} {\bibfield  {journal} {\bibinfo  {journal}
  {Phys. Rev. E}\ }\textbf {\bibinfo {volume} {97}},\ \bibinfo {pages} {042413}
  (\bibinfo {year} {2018})}\BibitemShut {NoStop}%
\bibitem [{\citenamefont {Okuda}\ and\ \citenamefont
  {Sato}(2022)}]{Okuda:2022a}%
  \BibitemOpen
  \bibfield  {author} {\bibinfo {author} {\bibfnamefont {S.}~\bibnamefont
  {Okuda}}\ and\ \bibinfo {author} {\bibfnamefont {K.}~\bibnamefont {Sato}},\
  }\href@noop {} {\bibfield  {journal} {\bibinfo  {journal} {Biophys. J.}\
  }\textbf {\bibinfo {volume} {121}},\ \bibinfo {pages} {1856} (\bibinfo {year}
  {2022})}\BibitemShut {NoStop}%
\bibitem [{\citenamefont {Matsushita}\ \emph {et~al.}(2022)\citenamefont
  {Matsushita}, \citenamefont {Hashimura}, \citenamefont {Kuwayama},\ and\
  \citenamefont {Fujimoto}}]{Matsushita:2022a}%
  \BibitemOpen
  \bibfield  {author} {\bibinfo {author} {\bibfnamefont {K.}~\bibnamefont
  {Matsushita}}, \bibinfo {author} {\bibfnamefont {H.}~\bibnamefont
  {Hashimura}}, \bibinfo {author} {\bibfnamefont {H.}~\bibnamefont
  {Kuwayama}},\ and\ \bibinfo {author} {\bibfnamefont {K.}~\bibnamefont
  {Fujimoto}},\ }\href@noop {} {\bibfield  {journal} {\bibinfo  {journal} {J.
  Phys. Soc. Jpn.}\ }\textbf {\bibinfo {volume} {91}},\ \bibinfo {pages}
  {054802} (\bibinfo {year} {2022})}\BibitemShut {NoStop}%
\bibitem [{\citenamefont {Okuda}\ \emph {et~al.}(2022)\citenamefont {Okuda},
  \citenamefont {Sato},\ and\ \citenamefont {Hiraiwa}}]{Okuda:2022b}%
  \BibitemOpen
  \bibfield  {author} {\bibinfo {author} {\bibfnamefont {S.}~\bibnamefont
  {Okuda}}, \bibinfo {author} {\bibfnamefont {K.}~\bibnamefont {Sato}},\ and\
  \bibinfo {author} {\bibfnamefont {T.}~\bibnamefont {Hiraiwa}},\ }\href@noop
  {} {\bibfield  {journal} {\bibinfo  {journal} {Eur. Phys. J. E}\ }\textbf
  {\bibinfo {volume} {45}},\ \bibinfo {pages} {69} (\bibinfo {year}
  {2022})}\BibitemShut {NoStop}%
\bibitem [{\citenamefont {Alert}\ and\ \citenamefont
  {Trepat}(2020)}]{Alert:2020}%
  \BibitemOpen
  \bibfield  {author} {\bibinfo {author} {\bibfnamefont {R.}~\bibnamefont
  {Alert}}\ and\ \bibinfo {author} {\bibfnamefont {X.}~\bibnamefont {Trepat}},\
  }\href@noop {} {\bibfield  {journal} {\bibinfo  {journal} {Annu. Rev.
  Condens. Matter Phys.}\ }\textbf {\bibinfo {volume} {11}},\ \bibinfo {pages}
  {77} (\bibinfo {year} {2020})}\BibitemShut {NoStop}%
\bibitem [{\citenamefont {Levan}(1981)}]{Levan:1981}%
  \BibitemOpen
  \bibfield  {author} {\bibinfo {author} {\bibfnamefont {M.~D.}\ \bibnamefont
  {Levan}},\ }\href@noop {} {\bibfield  {journal} {\bibinfo  {journal} {J.
  Colloid Interface Sci.}\ }\textbf {\bibinfo {volume} {83}},\ \bibinfo {pages}
  {11} (\bibinfo {year} {1981})}\BibitemShut {NoStop}%
\bibitem [{\citenamefont {Shellard}\ and\ \citenamefont
  {Mayor}(2019)}]{Shellard:2019}%
  \BibitemOpen
  \bibfield  {author} {\bibinfo {author} {\bibfnamefont {A.}~\bibnamefont
  {Shellard}}\ and\ \bibinfo {author} {\bibfnamefont {R.}~\bibnamefont
  {Mayor}},\ }\href@noop {} {\bibfield  {journal} {\bibinfo  {journal} {J.
  Cell. Sci}\ }\textbf {\bibinfo {volume} {132 (8)}},\ \bibinfo {pages}
  {jcs226142} (\bibinfo {year} {2019})}\BibitemShut {NoStop}%
\bibitem [{\citenamefont {Itatani}\ and\ \citenamefont
  {Nabika}(2022)}]{Itatani:2022}%
  \BibitemOpen
  \bibfield  {author} {\bibinfo {author} {\bibfnamefont {M.}~\bibnamefont
  {Itatani}}\ and\ \bibinfo {author} {\bibfnamefont {H.}~\bibnamefont
  {Nabika}},\ }\href@noop {} {\bibfield  {journal} {\bibinfo  {journal} {Front.
  Phys.}\ }\textbf {\bibinfo {volume} {10}},\ \bibinfo {pages} {849111}
  (\bibinfo {year} {2022})}\BibitemShut {NoStop}%
\bibitem [{\citenamefont {Yadav}\ \emph {et~al.}(2022)\citenamefont {Yadav},
  \citenamefont {Yousafzai}, \citenamefont {Amiri}, \citenamefont {Style},
  \citenamefont {Dufresne},\ and\ \citenamefont {Murrell}}]{Yadav:2022}%
  \BibitemOpen
  \bibfield  {author} {\bibinfo {author} {\bibfnamefont {V.}~\bibnamefont
  {Yadav}}, \bibinfo {author} {\bibfnamefont {M.~S.}\ \bibnamefont
  {Yousafzai}}, \bibinfo {author} {\bibfnamefont {S.}~\bibnamefont {Amiri}},
  \bibinfo {author} {\bibfnamefont {R.~W.}\ \bibnamefont {Style}}, \bibinfo
  {author} {\bibfnamefont {E.~R.}\ \bibnamefont {Dufresne}},\ and\ \bibinfo
  {author} {\bibfnamefont {M.}~\bibnamefont {Murrell}},\ }\href@noop {}
  {\bibfield  {journal} {\bibinfo  {journal} {Phys. Rev. Fluid}\ }\textbf
  {\bibinfo {volume} {7}},\ \bibinfo {pages} {L031101} (\bibinfo {year}
  {2022})}\BibitemShut {NoStop}%
\bibitem [{\citenamefont {Pajic-Lijakovic}\ \emph {et~al.}(2023)\citenamefont
  {Pajic-Lijakovic}, \citenamefont {Eftimie}, \citenamefont {Milivojevic},\
  and\ \citenamefont {Bordas}}]{Pajic-Lijakovic:2023}%
  \BibitemOpen
  \bibfield  {author} {\bibinfo {author} {\bibfnamefont {I.}~\bibnamefont
  {Pajic-Lijakovic}}, \bibinfo {author} {\bibfnamefont {R.}~\bibnamefont
  {Eftimie}}, \bibinfo {author} {\bibfnamefont {M.}~\bibnamefont
  {Milivojevic}},\ and\ \bibinfo {author} {\bibfnamefont {S.~P.}\ \bibnamefont
  {Bordas}},\ }\href@noop {} {\bibfield  {journal} {\bibinfo  {journal} {Semin.
  Cell Dev. Biol.}\ }\textbf {\bibinfo {volume} {147}},\ \bibinfo {pages} {34}
  (\bibinfo {year} {2023})}\BibitemShut {NoStop}%
\bibitem [{\citenamefont {Sato}(2023)}]{Sato:2023}%
  \BibitemOpen
  \bibfield  {author} {\bibinfo {author} {\bibfnamefont {K.}~\bibnamefont
  {Sato}},\ }\href@noop {} {\bibfield  {journal} {\bibinfo  {journal} {Front.
  Cell Dev. Biol.}\ }\textbf {\bibinfo {volume} {11}},\ \bibinfo {pages}
  {1126819} (\bibinfo {year} {2023})}\BibitemShut {NoStop}%
\bibitem [{\citenamefont {Sesaki}\ and\ \citenamefont
  {Siu}(1996)}]{Sesaki:1996}%
  \BibitemOpen
  \bibfield  {author} {\bibinfo {author} {\bibfnamefont {H.}~\bibnamefont
  {Sesaki}}\ and\ \bibinfo {author} {\bibfnamefont {C.-H.}\ \bibnamefont
  {Siu}},\ }\href@noop {} {\bibfield  {journal} {\bibinfo  {journal} {Dev.
  Biol.}\ }\textbf {\bibinfo {volume} {177}},\ \bibinfo {pages} {504} (\bibinfo
  {year} {1996})}\BibitemShut {NoStop}%
\bibitem [{\citenamefont {Matsushita}(2017)}]{Matsushita:2017}%
  \BibitemOpen
  \bibfield  {author} {\bibinfo {author} {\bibfnamefont {K.}~\bibnamefont
  {Matsushita}},\ }\href@noop {} {\bibfield  {journal} {\bibinfo  {journal}
  {Phys. Rev. E}\ }\textbf {\bibinfo {volume} {95}},\ \bibinfo {pages} {032415}
  (\bibinfo {year} {2017})}\BibitemShut {NoStop}%
\bibitem [{\citenamefont {Matsushita}\ \emph {et~al.}(2021)\citenamefont
  {Matsushita}, \citenamefont {Yabunaka},\ and\ \citenamefont
  {Fujimoto}}]{Matsushita:2021a}%
  \BibitemOpen
  \bibfield  {author} {\bibinfo {author} {\bibfnamefont {K.}~\bibnamefont
  {Matsushita}}, \bibinfo {author} {\bibfnamefont {S.}~\bibnamefont
  {Yabunaka}},\ and\ \bibinfo {author} {\bibfnamefont {K.}~\bibnamefont
  {Fujimoto}},\ }\href@noop {} {\bibfield  {journal} {\bibinfo  {journal} {J.
  Phys. Soc. Jpn.}\ }\textbf {\bibinfo {volume} {90}},\ \bibinfo {pages}
  {054801} (\bibinfo {year} {2021})}\BibitemShut {NoStop}%
\bibitem [{\citenamefont {Weber}\ \emph {et~al.}(2013)\citenamefont {Weber},
  \citenamefont {Hanke}, \citenamefont {Deseigne}, \citenamefont {L\'{e}onard},
  \citenamefont {Dauchot}, \citenamefont {Frey},\ and\ \citenamefont
  {Chat\'{e}}}]{Weber:2013}%
  \BibitemOpen
  \bibfield  {author} {\bibinfo {author} {\bibfnamefont {C.~A.}\ \bibnamefont
  {Weber}}, \bibinfo {author} {\bibfnamefont {T.}~\bibnamefont {Hanke}},
  \bibinfo {author} {\bibfnamefont {J.}~\bibnamefont {Deseigne}}, \bibinfo
  {author} {\bibfnamefont {S.}~\bibnamefont {L\'{e}onard}}, \bibinfo {author}
  {\bibfnamefont {O.}~\bibnamefont {Dauchot}}, \bibinfo {author} {\bibfnamefont
  {E.}~\bibnamefont {Frey}},\ and\ \bibinfo {author} {\bibfnamefont
  {H.}~\bibnamefont {Chat\'{e}}},\ }\href@noop {} {\bibfield  {journal}
  {\bibinfo  {journal} {Phys. Rev. Lett.}\ }\textbf {\bibinfo {volume} {110}},\
  \bibinfo {pages} {208001} (\bibinfo {year} {2013})}\BibitemShut {NoStop}%
\bibitem [{\citenamefont {Hanke}\ \emph {et~al.}(2013)\citenamefont {Hanke},
  \citenamefont {Weber},\ and\ \citenamefont {Frey}}]{Hanke:2013}%
  \BibitemOpen
  \bibfield  {author} {\bibinfo {author} {\bibfnamefont {T.}~\bibnamefont
  {Hanke}}, \bibinfo {author} {\bibfnamefont {C.~A.}\ \bibnamefont {Weber}},\
  and\ \bibinfo {author} {\bibfnamefont {E.}~\bibnamefont {Frey}},\ }\href@noop
  {} {\bibfield  {journal} {\bibinfo  {journal} {Phys. Rev. E}\ }\textbf
  {\bibinfo {volume} {88}},\ \bibinfo {pages} {052309} (\bibinfo {year}
  {2013})}\BibitemShut {NoStop}%
\bibitem [{\citenamefont {Hiraoka}\ \emph {et~al.}(2016)\citenamefont
  {Hiraoka}, \citenamefont {Shimada},\ and\ \citenamefont
  {Ito}}]{Hiraoka:2016}%
  \BibitemOpen
  \bibfield  {author} {\bibinfo {author} {\bibfnamefont {T.}~\bibnamefont
  {Hiraoka}}, \bibinfo {author} {\bibfnamefont {T.}~\bibnamefont {Shimada}},\
  and\ \bibinfo {author} {\bibfnamefont {N.}~\bibnamefont {Ito}},\ }\href@noop
  {} {\bibfield  {journal} {\bibinfo  {journal} {Phys. Rev. E}\ }\textbf
  {\bibinfo {volume} {94}},\ \bibinfo {pages} {062612} (\bibinfo {year}
  {2016})}\BibitemShut {NoStop}%
\bibitem [{\citenamefont {Hiraoka}\ \emph {et~al.}(2017)\citenamefont
  {Hiraoka}, \citenamefont {Shimada},\ and\ \citenamefont
  {Ito}}]{Hiraoka:2017}%
  \BibitemOpen
  \bibfield  {author} {\bibinfo {author} {\bibfnamefont {T.}~\bibnamefont
  {Hiraoka}}, \bibinfo {author} {\bibfnamefont {T.}~\bibnamefont {Shimada}},\
  and\ \bibinfo {author} {\bibfnamefont {N.}~\bibnamefont {Ito}},\ }\href@noop
  {} {\bibfield  {journal} {\bibinfo  {journal} {J. Phys: Conf. Series}\
  }\textbf {\bibinfo {volume} {921}},\ \bibinfo {pages} {012006} (\bibinfo
  {year} {2017})}\BibitemShut {NoStop}%
\bibitem [{\citenamefont {Matsushita}\ \emph {et~al.}(2023)\citenamefont
  {Matsushita}, \citenamefont {Arakaki}, \citenamefont {Kamamoto},
  \citenamefont {Sudo},\ and\ \citenamefont {Fujimoto}}]{Matsushita:2023a}%
  \BibitemOpen
  \bibfield  {author} {\bibinfo {author} {\bibfnamefont {K.}~\bibnamefont
  {Matsushita}}, \bibinfo {author} {\bibfnamefont {T.}~\bibnamefont {Arakaki}},
  \bibinfo {author} {\bibfnamefont {N.}~\bibnamefont {Kamamoto}}, \bibinfo
  {author} {\bibfnamefont {M.}~\bibnamefont {Sudo}},\ and\ \bibinfo {author}
  {\bibfnamefont {K.}~\bibnamefont {Fujimoto}},\ }\href@noop {} {\bibfield
  {journal} {\bibinfo  {journal} {Sympo. Traffic Flow Self-driven Particles}\
  }\textbf {\bibinfo {volume} {28}},\ \bibinfo {pages} {5} (\bibinfo {year}
  {2023})}\BibitemShut {NoStop}%
\bibitem [{\citenamefont {Matsumoto}\ \emph {et~al.}(2024)\citenamefont
  {Matsumoto}, \citenamefont {Matsushita}, \citenamefont {Hane}, \citenamefont
  {Wen}, \citenamefont {Kurematsu}, \citenamefont {Ota}, \citenamefont
  {Nguyen}, \citenamefont {Thai}, \citenamefont {Herranz-P\'{e}rez},
  \citenamefont {Sawada}, \citenamefont {Fujimoto}, \citenamefont
  {García-Verdugo}, \citenamefont {Kimura}, \citenamefont {Seki},
  \citenamefont {Sato}, \citenamefont {Ohno},\ and\ \citenamefont
  {Sawamoto}}]{Matsumoto:2024}%
  \BibitemOpen
  \bibfield  {author} {\bibinfo {author} {\bibfnamefont {M.}~\bibnamefont
  {Matsumoto}}, \bibinfo {author} {\bibfnamefont {K.}~\bibnamefont
  {Matsushita}}, \bibinfo {author} {\bibfnamefont {M.}~\bibnamefont {Hane}},
  \bibinfo {author} {\bibfnamefont {C.}~\bibnamefont {Wen}}, \bibinfo {author}
  {\bibfnamefont {C.}~\bibnamefont {Kurematsu}}, \bibinfo {author}
  {\bibfnamefont {H.}~\bibnamefont {Ota}}, \bibinfo {author} {\bibfnamefont
  {H.~B.}\ \bibnamefont {Nguyen}}, \bibinfo {author} {\bibfnamefont {T.~Q.}\
  \bibnamefont {Thai}}, \bibinfo {author} {\bibfnamefont {V.}~\bibnamefont
  {Herranz-P\'{e}rez}}, \bibinfo {author} {\bibfnamefont {M.}~\bibnamefont
  {Sawada}}, \bibinfo {author} {\bibfnamefont {K.}~\bibnamefont {Fujimoto}},
  \bibinfo {author} {\bibfnamefont {J.~M.}\ \bibnamefont {García-Verdugo}},
  \bibinfo {author} {\bibfnamefont {K.~D.}\ \bibnamefont {Kimura}}, \bibinfo
  {author} {\bibfnamefont {T.}~\bibnamefont {Seki}}, \bibinfo {author}
  {\bibfnamefont {C.}~\bibnamefont {Sato}}, \bibinfo {author} {\bibfnamefont
  {N.}~\bibnamefont {Ohno}},\ and\ \bibinfo {author} {\bibfnamefont
  {K.}~\bibnamefont {Sawamoto}},\ }\href@noop {} {\bibfield  {journal}
  {\bibinfo  {journal} {EMBO Mol. Med.}\ }\textbf {\bibinfo {volume} {16}},\
  \bibinfo {pages} {1228} (\bibinfo {year} {2024})}\BibitemShut {NoStop}%
\bibitem [{\citenamefont {Hawkes}\ and\ \citenamefont
  {Wang}(1982)}]{Hawkes:1982}%
  \BibitemOpen
  \bibfield  {author} {\bibinfo {author} {\bibfnamefont {S.}~\bibnamefont
  {Hawkes}}\ and\ \bibinfo {author} {\bibfnamefont {J.~L.}\ \bibnamefont
  {Wang}},\ }\href@noop {} {\emph {\bibinfo {title} {Extracellular Matrix}}}\
  (\bibinfo  {publisher} {Academic Press, London},\ \bibinfo {year}
  {1982})\BibitemShut {NoStop}%
\bibitem [{\citenamefont {Mecham}(2011)}]{Mecham:2011}%
  \BibitemOpen
  \bibfield  {author} {\bibinfo {author} {\bibfnamefont {R.}~\bibnamefont
  {Mecham}},\ }\href@noop {} {\emph {\bibinfo {title} {The Extracellular
  Matrix: an Overview}}}\ (\bibinfo  {publisher} {Springer-Verlag, Berlin
  Heidelberg},\ \bibinfo {year} {2011})\BibitemShut {NoStop}%
\bibitem [{\citenamefont {Graner}\ and\ \citenamefont
  {Glazier}(1992)}]{Graner:1992}%
  \BibitemOpen
  \bibfield  {author} {\bibinfo {author} {\bibfnamefont {F.}~\bibnamefont
  {Graner}}\ and\ \bibinfo {author} {\bibfnamefont {J.~A.}\ \bibnamefont
  {Glazier}},\ }\href@noop {} {\bibfield  {journal} {\bibinfo  {journal}
  {Phys.~Rev.~Lett.}\ }\textbf {\bibinfo {volume} {69}},\ \bibinfo {pages}
  {2013} (\bibinfo {year} {1992})}\BibitemShut {NoStop}%
\bibitem [{\citenamefont {Glazier}\ and\ \citenamefont
  {Graner}(1993)}]{Glazier:1993}%
  \BibitemOpen
  \bibfield  {author} {\bibinfo {author} {\bibfnamefont {J.~A.}\ \bibnamefont
  {Glazier}}\ and\ \bibinfo {author} {\bibfnamefont {F.}~\bibnamefont
  {Graner}},\ }\href@noop {} {\bibfield  {journal} {\bibinfo  {journal} {Phys.
  Rev. E}\ }\textbf {\bibinfo {volume} {47}},\ \bibinfo {pages} {2128}
  (\bibinfo {year} {1993})}\BibitemShut {NoStop}%
\bibitem [{\citenamefont {Graner}(1993)}]{Graner:1993}%
  \BibitemOpen
  \bibfield  {author} {\bibinfo {author} {\bibfnamefont {F.}~\bibnamefont
  {Graner}},\ }\href@noop {} {\bibfield  {journal} {\bibinfo  {journal} {J.
  Theor. Biol.}\ }\textbf {\bibinfo {volume} {164}},\ \bibinfo {pages} {455}
  (\bibinfo {year} {1993})}\BibitemShut {NoStop}%
\bibitem [{\citenamefont {Anderson}\ \emph {et~al.}(2007)\citenamefont
  {Anderson}, \citenamefont {Chaplain},\ and\ \citenamefont
  {Rejniak}}]{Anderson:2007}%
  \BibitemOpen
  \bibfield  {author} {\bibinfo {author} {\bibfnamefont {A.~R.~A.}\
  \bibnamefont {Anderson}}, \bibinfo {author} {\bibfnamefont {M.~A.~J.}\
  \bibnamefont {Chaplain}},\ and\ \bibinfo {author} {\bibfnamefont {K.~A.}\
  \bibnamefont {Rejniak}},\ }\href@noop {} {\emph {\bibinfo {title}
  {Single-Cell-Based Models in Biology and Medicine}}}\ (\bibinfo  {publisher}
  {Birkhauser Verlag AG, Basel},\ \bibinfo {year} {2007})\BibitemShut {NoStop}%
\bibitem [{\citenamefont {Scianna}\ and\ \citenamefont
  {Preziosi}(2013)}]{Scianna:2013}%
  \BibitemOpen
  \bibfield  {author} {\bibinfo {author} {\bibfnamefont {M.}~\bibnamefont
  {Scianna}}\ and\ \bibinfo {author} {\bibfnamefont {L.}~\bibnamefont
  {Preziosi}},\ }\href@noop {} {\emph {\bibinfo {title} {Cellular Potts
  Model}}}\ (\bibinfo  {publisher} {CRC Press, UK},\ \bibinfo {year}
  {2013})\BibitemShut {NoStop}%
\bibitem [{\citenamefont {Hirashima}\ \emph {et~al.}(2017)\citenamefont
  {Hirashima}, \citenamefont {Rens},\ and\ \citenamefont
  {Merks}}]{Hirashima:2017}%
  \BibitemOpen
  \bibfield  {author} {\bibinfo {author} {\bibfnamefont {T.}~\bibnamefont
  {Hirashima}}, \bibinfo {author} {\bibfnamefont {E.~G.}\ \bibnamefont
  {Rens}},\ and\ \bibinfo {author} {\bibfnamefont {R.~M.~H.}\ \bibnamefont
  {Merks}},\ }\href@noop {} {\bibfield  {journal} {\bibinfo  {journal} {Dev.
  Growth Differ.}\ }\textbf {\bibinfo {volume} {59}},\ \bibinfo {pages} {329}
  (\bibinfo {year} {2017})}\BibitemShut {NoStop}%
\bibitem [{\citenamefont {Clifford}\ and\ \citenamefont
  {Sudbury}(1973)}]{Clifford:1973}%
  \BibitemOpen
  \bibfield  {author} {\bibinfo {author} {\bibfnamefont {P.}~\bibnamefont
  {Clifford}}\ and\ \bibinfo {author} {\bibfnamefont {A.}~\bibnamefont
  {Sudbury}},\ }\href@noop {} {\bibfield  {journal} {\bibinfo  {journal}
  {Biometrica}\ }\textbf {\bibinfo {volume} {60}},\ \bibinfo {pages} {581}
  (\bibinfo {year} {1973})}\BibitemShut {NoStop}%
\bibitem [{\citenamefont {Liggett}(1985)}]{Liggett:1985}%
  \BibitemOpen
  \bibfield  {author} {\bibinfo {author} {\bibfnamefont {L.~M.}\ \bibnamefont
  {Liggett}},\ }\href@noop {} {\emph {\bibinfo {title} {Interacting Particle
  Systems}}}\ (\bibinfo  {publisher} {Springer, New York},\ \bibinfo {year}
  {1985})\BibitemShut {NoStop}%
\bibitem [{\citenamefont {Landau}\ and\ \citenamefont
  {Binder}(2021)}]{Landau:2021}%
  \BibitemOpen
  \bibfield  {author} {\bibinfo {author} {\bibfnamefont {D.}~\bibnamefont
  {Landau}}\ and\ \bibinfo {author} {\bibfnamefont {K.}~\bibnamefont
  {Binder}},\ }\href@noop {} {\emph {\bibinfo {title} {A Guide to Monte Carlo
  Simulations in Statistical Physics}}}\ (\bibinfo  {publisher} {Cambridge
  University Press, Cambridge},\ \bibinfo {year} {2021})\BibitemShut {NoStop}%
\bibitem [{\citenamefont {de~Gennes}\ \emph {et~al.}(2004)\citenamefont
  {de~Gennes}, \citenamefont {Brochard-Wyart},\ and\ \citenamefont
  {Quere}}]{DeGennes:2004}%
  \BibitemOpen
  \bibfield  {author} {\bibinfo {author} {\bibfnamefont {P.-G.}\ \bibnamefont
  {de~Gennes}}, \bibinfo {author} {\bibfnamefont {F.}~\bibnamefont
  {Brochard-Wyart}},\ and\ \bibinfo {author} {\bibfnamefont {D.}~\bibnamefont
  {Quere}},\ }\href@noop {} {\emph {\bibinfo {title} {Capillarity and Wetting
  Phenomena: Drops, Bubbles, Pearls, Waves}}}\ (\bibinfo  {publisher}
  {Springer},\ \bibinfo {year} {2004})\BibitemShut {NoStop}%
\bibitem [{\citenamefont {Schwarz}\ and\ \citenamefont
  {Safran}(2013)}]{Safran:2013}%
  \BibitemOpen
  \bibfield  {author} {\bibinfo {author} {\bibfnamefont {U.~S.}\ \bibnamefont
  {Schwarz}}\ and\ \bibinfo {author} {\bibfnamefont {S.~A.}\ \bibnamefont
  {Safran}},\ }\href@noop {} {\bibfield  {journal} {\bibinfo  {journal} {Rev.
  Mod. Phys.}\ }\textbf {\bibinfo {volume} {85}},\ \bibinfo {pages} {1327}
  (\bibinfo {year} {2013})}\BibitemShut {NoStop}%
\bibitem [{\citenamefont {Takeichi}(2014)}]{Takeichi:2014}%
  \BibitemOpen
  \bibfield  {author} {\bibinfo {author} {\bibfnamefont {M.}~\bibnamefont
  {Takeichi}},\ }\href@noop {} {\bibfield  {journal} {\bibinfo  {journal} {Nat.
  Rev. Mol. Cell. Biol.}\ }\textbf {\bibinfo {volume} {15}},\ \bibinfo {pages}
  {397} (\bibinfo {year} {2014})}\BibitemShut {NoStop}%
\bibitem [{\citenamefont {Zajac}\ \emph {et~al.}(2002)\citenamefont {Zajac},
  \citenamefont {Jonesa},\ and\ \citenamefont {Glazier}}]{Zajac:2002}%
  \BibitemOpen
  \bibfield  {author} {\bibinfo {author} {\bibfnamefont {M.}~\bibnamefont
  {Zajac}}, \bibinfo {author} {\bibfnamefont {G.~L.}\ \bibnamefont {Jonesa}},\
  and\ \bibinfo {author} {\bibfnamefont {J.~A.}\ \bibnamefont {Glazier}},\
  }\href@noop {} {\bibfield  {journal} {\bibinfo  {journal} {J. Theor. Biol.}\
  }\textbf {\bibinfo {volume} {222}},\ \bibinfo {pages} {247} (\bibinfo {year}
  {2002})}\BibitemShut {NoStop}%
\bibitem [{\citenamefont {Vroomans}\ \emph {et~al.}(2015)\citenamefont
  {Vroomans}, \citenamefont {Hogeweg},\ and\ \citenamefont {ten
  Tusscher}}]{Vroomans:2015}%
  \BibitemOpen
  \bibfield  {author} {\bibinfo {author} {\bibfnamefont {R.~M.~A.}\
  \bibnamefont {Vroomans}}, \bibinfo {author} {\bibfnamefont {P.}~\bibnamefont
  {Hogeweg}},\ and\ \bibinfo {author} {\bibfnamefont {K.~H. W.~J.}\
  \bibnamefont {ten Tusscher}},\ }\href@noop {} {\bibfield  {journal} {\bibinfo
   {journal} {PLoS Comput. Biol.}\ }\textbf {\bibinfo {volume} {11}},\ \bibinfo
  {pages} {e1004092} (\bibinfo {year} {2015})}\BibitemShut {NoStop}%
\bibitem [{\citenamefont {Bi}\ \emph {et~al.}(2016)\citenamefont {Bi},
  \citenamefont {Yang}, \citenamefont {Marchetti},\ and\ \citenamefont
  {Manning}}]{Bi:2016}%
  \BibitemOpen
  \bibfield  {author} {\bibinfo {author} {\bibfnamefont {D.}~\bibnamefont
  {Bi}}, \bibinfo {author} {\bibfnamefont {X.}~\bibnamefont {Yang}}, \bibinfo
  {author} {\bibfnamefont {M.~C.}\ \bibnamefont {Marchetti}},\ and\ \bibinfo
  {author} {\bibfnamefont {M.~L.}\ \bibnamefont {Manning}},\ }\href@noop {}
  {\bibfield  {journal} {\bibinfo  {journal} {Phys. Rev. X}\ }\textbf {\bibinfo
  {volume} {6}},\ \bibinfo {pages} {021011} (\bibinfo {year}
  {2016})}\BibitemShut {NoStop}%
\bibitem [{\citenamefont {Pinheiro}\ and\ \citenamefont
  {Mitchel}(2024)}]{Pinheiro:2024}%
  \BibitemOpen
  \bibfield  {author} {\bibinfo {author} {\bibfnamefont {D.}~\bibnamefont
  {Pinheiro}}\ and\ \bibinfo {author} {\bibfnamefont {J.}~\bibnamefont
  {Mitchel}},\ }\href@noop {} {\bibfield  {journal} {\bibinfo  {journal} {Curr.
  Op. Cell Biol.}\ }\textbf {\bibinfo {volume} {86}},\ \bibinfo {pages}
  {102310} (\bibinfo {year} {2024})}\BibitemShut {NoStop}%
\bibitem [{\citenamefont {Beug}\ \emph {et~al.}(1973)\citenamefont {Beug},
  \citenamefont {Katz},\ and\ \citenamefont {Gerisch}}]{Beug:1973}%
  \BibitemOpen
  \bibfield  {author} {\bibinfo {author} {\bibfnamefont {H.}~\bibnamefont
  {Beug}}, \bibinfo {author} {\bibfnamefont {F.~E.}\ \bibnamefont {Katz}},\
  and\ \bibinfo {author} {\bibfnamefont {G.}~\bibnamefont {Gerisch}},\
  }\href@noop {} {\bibfield  {journal} {\bibinfo  {journal} {J. Cell Biol.}\
  }\textbf {\bibinfo {volume} {56}},\ \bibinfo {pages} {647} (\bibinfo {year}
  {1973})}\BibitemShut {NoStop}%
\bibitem [{\citenamefont {M\"{u}ller}\ and\ \citenamefont
  {Gerisch}(1978)}]{Muller:1978}%
  \BibitemOpen
  \bibfield  {author} {\bibinfo {author} {\bibfnamefont {K.}~\bibnamefont
  {M\"{u}ller}}\ and\ \bibinfo {author} {\bibfnamefont {G.}~\bibnamefont
  {Gerisch}},\ }\href@noop {} {\bibfield  {journal} {\bibinfo  {journal}
  {Nature (London)}\ }\textbf {\bibinfo {volume} {274}},\ \bibinfo {pages}
  {445} (\bibinfo {year} {1978})}\BibitemShut {NoStop}%
\end{thebibliography}
\end{document}